\documentclass[fleqn,usenatbib]{mnras}

\usepackage{newtxtext,newtxmath}

\usepackage[T1]{fontenc}

\DeclareRobustCommand{\VAN}[3]{#2}
\let\VANthebibliography\thebibliography
\def\thebibliography{\DeclareRobustCommand{\VAN}[3]{##3}\VANthebibliography}

\usepackage{graphicx}	
\usepackage{amsmath}	
\usepackage{xcolor}
\usepackage{mwe}
\usepackage{subfig}
\usepackage[normalem]{ulem}
\usepackage{tabularx}
\usepackage[roman]{parnotes}
\usepackage{mathtools}
\usepackage{multirow}
\usepackage[thinc]{esdiff}
\usepackage{physics}
\usepackage{cleveref}
\usepackage{tabularray}
\usepackage{mathtools}

\renewcommand*\vec[1]{\ensuremath{\boldsymbol{#1}}}

\title[Clustering from voids and galaxies in SDSS]{Cosmological measurements from void-galaxy and galaxy-galaxy clustering in the Sloan Digital Sky Survey}

\author[A. Woodfinden et al.]{Alex Woodfinden,$^{1,2}$\thanks{E-mail: awoodfin@uwaterloo.ca} Will J. Percival,$^{1,2,3}$ Seshadri Nadathur,$^{4}$ Hans A. Winther,$^{5}$ T. S. Fraser,$^{1,2}$ \newauthor{Elena Massara,$^{1,2}$ Enrique Paillas,$^{1,2}$ Slađana Radinović$^{5}$}
\\
$^{1}$Waterloo Centre for Astrophysics, University of Waterloo, 200 University Ave W, Waterloo, ON N2L 3G1, Canada\\
$^{2}$Department of Physics and Astronomy, University of Waterloo, 200 University Ave W, Waterloo, ON N2L 3G1, Canada\\
$^{3}$Perimeter Institute for Theoretical Physics, 31 Caroline St North, Waterloo, ON N2L 2Y5, Canada\\
$^{4}$Institute of Cosmology and Gravitation, University of Portsmouth, Burnaby Road, Portsmouth, PO1 3FX, United Kingdom\\
$^{5}$Institute of Theoretical Astrophysics, University of Oslo, P.O. Box 1029 Blindern, N-0315 Oslo, Norway
}

\date{Accepted XXX. Received YYY; in original form ZZZ}

\pubyear{2022}

\begin{document}
\label{firstpage}
\pagerange{\pageref{firstpage}--\pageref{lastpage}}
\maketitle

\begin{abstract}
We present the cosmological implications of measurements of void-galaxy and galaxy-galaxy clustering from the Sloan Digital Sky Survey (SDSS)  Main Galaxy Sample (MGS), Baryon Oscillation Spectroscopic Survey (BOSS), and extended BOSS (eBOSS) luminous red galaxy catalogues from SDSS Data Release 7, 12, and 16, covering the redshift range $0.07 < z < 1.0$. We fit a standard $\Lambda$CDM cosmological model as well as various extensions including a constant dark energy equation of state not equal to $-1$, a time-varying dark energy equation of state, and these same models allowing for spatial curvature. Results on key parameters of these models are reported for void-galaxy and galaxy-galaxy clustering alone, both of these combined, and all these combined with measurements from the cosmic microwave background (CMB) and supernovae (SN). For the combination of void-galaxy and galaxy-galaxy clustering, we find tight constraints of $\Omega_\mathrm{m} = 0.356\pm 0.024$ for a base $\Lambda$CDM cosmology, $\Omega_\mathrm{m} = 0.391^{+0.028}_{-0.021}, w = -1.50^{+0.43}_{-0.28}$ additionally allowing the dark energy equation of state $w$ to vary, and $\Omega_\mathrm{m} = 0.331^{+0.067}_{-0.094}, w=-1.41^{+0.70}_{-0.31},\ \mathrm{and}\ \Omega_\mathrm{k} = 0.06^{+0.18}_{-0.13}$ further extending to non-flat models. The combined SDSS results from void-galaxy and galaxy-galaxy clustering in combination with CMB+SN provide a 30\% improvement in parameter $\Omega_\mathrm{m}$ over CMB+SN for $\Lambda$CDM,  a 5\% improvement in parameter $\Omega_\mathrm{m}$ when $w$ is allowed to vary, and a 32\% and 68\% improvement in parameters $\Omega_\mathrm{m}$ and $\Omega_\mathrm{k}$ when allowing for spatial curvature.

\end{abstract}




\section{Introduction}

The large-scale structure of the Universe contains a wealth of information about the expansion history of the Universe as well as the growth of structure within it. Measurements of these from spectroscopic galaxy surveys within the low-redshift Universe, in combination with Cosmic Microwave Background observations from \citet{Planck:2020}, provide the best evidence currently available for the standard $\Lambda$ Cold Dark Matter ($\Lambda$CDM) cosmological model. Modern spectroscopic galaxy surveys are focused on observing the baryon acoustic oscillation (BAO) feature, a relic of primordial sound waves, which can be used as a standard ruler \citep[e.g.][]{Alam-DR11&12:2015,Alam-eBOSS:2021}. In addition to the BAO feature, cosmological information can be extracted using various other techniques including redshift-space distortions \citep[RSD, ][]{Kaiser:1987} and the galaxy distribution around voids \citep{Lavaux:2012}. This work combines these three cosmological measurements from the same surveys to provide powerful constraints on the $\Lambda$CDM cosmological model and various extensions to it. 

Although the Universe is expected to be statistically homogeneous and isotropic on large scales, observations typically use different methods along and across the line of sight (LOS). For example, galaxy surveys measure the angular positions of galaxies across the LOS and redshifts along it. Consequently, the information provided by the observations differs with the angle to the LOS: In particular, separations along the LOS are sensitive to $D_\mathrm{H}(z) \equiv c/H(z)$ while across the LOS separations are sensitive to the comoving angular diameter distance $D_\mathrm{M}(z)$,
\begin{equation}\label{eq:DM}
    D_\mathrm{M}(z) = D_\mathrm{H}(0) \frac{1}{\sqrt{\Omega_\mathrm{k}}} \sinh{\left(\sqrt{\Omega_\mathrm{k}} \frac{D_\mathrm{C}}{D_\mathrm{H}(0)}\right)}
\end{equation} 
where 
\begin{equation}\label{eq:DC}
D_\mathrm{C}(z) \equiv \int_0^z dz' c / H(z')    
\end{equation} 
is the line-of-sight comoving distance, $H(z)$ is the Hubble expansion rate at redshift $z$, and $D_\mathrm{M}(z)$ is the transverse comoving distance to redshift $z$. To recover an isotropic map, we require $D_\mathrm{M}(z)/D_\mathrm{H}(z)$ to be correct when the redshift to distance conversion is done (this is known as the Alcock-Paczynski (AP) test, \citealt{Alcock:1979}). 

Measurements of the BAO feature bring in an extra dependence on $r_\mathrm{d}$, the comoving sound horizon at the baryon drag epoch, upon which the comoving BAO position depends. BAO observations with respect to the LOS allow us to perform geometrical measurements of both $D_\mathrm{H}(z)/r_\mathrm{d}$ (the LOS cosmological dependence of the BAO position) and $D_\mathrm{M}(z)/r_\mathrm{d}$ (the perpendicular cosmological dependence of the BAO position). Anisotropic BAO measurements thus intrinsically also include the Alcock-Paczynski test. 

Further complicating this picture are RSD, which cause additional anisotropic distortions. RSD are an artificial result of the peculiar velocities of galaxies caused by the growth of large-scale structure on the observed redshifts of galaxies \citep{Kaiser:1987}. When we incorrectly assume that the total redshift resulted from Hubble expansion, we imprint coherent anisotropies in the galaxy map. The amplitude of the large-scale RSD signal depends on $f(z)\sigma_8(z)$, where $f$ is the logarithmic growth rate of density perturbations and $\sigma_8$ is the amplitude of density fluctuations normalized using the standard deviation of density fluctuations in a sphere of $8 h^{-1}$ Mpc. Because the AP and RSD effects both give rise to anisotropic distortions, there is typically a degeneracy between measurements of both, such that combined observations of both BAO and RSD provide covariant measurements of $f(z)\sigma_8(z)$, $D_\mathrm{M}(z)/r_\mathrm{d}$ and $D_\mathrm{H}(z)/r_\mathrm{d}$ (hereafter we do not explicitly include the redshift dependence on $f\sigma_8$, $D_M$, or $D_H$). 

For a featureless power spectrum (i.e. a power law), there would exist a perfect degeneracy between $f\sigma_8$ and $D_\mathrm{M}/D_\mathrm{H}$. This degeneracy is broken by features in the power spectrum such as the BAO feature on large scales. However, the sample variance is significantly reduced if we can work on small scales. Unfortunately for the galaxy-galaxy auto-power, there are few features on small scales, and the degeneracy can only be mildly broken \citep{Ballinger:1996}. On the other hand, the small-scale void-galaxy cross-correlation has a number of features, and this degeneracy can be better broken, enhancing measurements.

The key to understanding how information can be extracted from void-galaxy cross-correlations is to consider a stack of voids. While voids do not have a known size and so cannot be easily used as a standard ruler like BAO, each void will have no preferred orientation in an isotropic Universe. Taking many voids and stacking the positions of galaxies in and around these voids should produce an apparent distribution that has spherical symmetry if the AP parameter $D_\mathrm{M}/D_\mathrm{H}$ is correct when the galaxy redshifts are converted to distances and no anisotropic selection bias is present in the creation of the void sample. Although voids have the potential to provide accurate measurements, there is an increased reliance on modelling non-linear astrophysical processes compared with BAO measurements potentially leading to increased systematic errors \citep[e.g.][]{Paz:2013,Hamaus:2016,Hawken:2017,Nadathur:2019c,Achitouv:2019,Hawken:2020,Aubert:2022a,Woodfinden:2022}.

The RSD contribution to anisotropy in voids can be modelled using linear theory with reasonable success and without needing to exclude small scales \citep{Hamaus:2014a,Cai:2016a,Nadathur:2019a,Paillas:2021}. This provides our constraints on $f\sigma_8$, however, with the caveat that the potential for systematic bias is higher due to the sparsity of tracers around void centres leading to a mismatch in the velocity profile of dark matter and tracers \citep{Massara:2022b}. The anisotropies in the void-galaxy cross-correlation function from RSD can be easily distinguished from those arising from the AP effect, and so measurements of the cross-correlation can be used to measure the AP parameter \citep{Hamaus:2015,Nadathur:2019c}. Work in \citet{Hamaus:2016,Nadathur:2019c,Nadathur:2020b} and \citet{Woodfinden:2022} applied this method to the BOSS and eBOSS surveys and resulted in a factor of between $1.7$ to $3.5$ more precise constraints than those from galaxy clustering and BAO in the same data. 

The best current constraints on cosmological models from galaxy spectroscopic surveys make use of the Sloan Digital Sky Survey (SDSS). Spectroscopic galaxy surveys were undertaken as part of the SDSS I, II, III and IV experiments \citep{York:2000,Eisenstein:2011,Blanton:2017}. For SDSS-III and SDSS-IV the galaxy surveys were called BOSS and eBOSS (these are described in greater detail in Section~\ref{sec:data}). In this work, we use all available SDSS data for luminous red galaxies (LRG). We take measurements of the distribution of galaxies around voids using the void-galaxy correlation function (hereon referred to as voids) \citep{Woodfinden:2022,Nadathur:2020b} as well as information from BAO and RSD from the same galaxy sample (hereon referred to as galaxies). We aim to combine this information to provide the most accurate cosmological constraints available from SDSS. 

Our paper is structured as follows: in Section~\ref{sec:data} we summarize the data and mock catalogues used to determine the measurements used in this work as well as determine their cross-covariance. In Section~\ref{sec:Combination} we review the theoretical background allowing consensus results between the void-galaxy and galaxy-galaxy clustering techniques using the same data set. In Section~\ref{sec:Results} we present the results of our analysis for a wide range of varying cosmologies and discuss the implications of this. Finally, we conclude in Section~\ref{sec:conclusions}. 

\section{Data and Mocks} \label{sec:data}
We use previous analyses of the void-galaxy correlation function published in \citet{Nadathur:2020b} and \citet{Woodfinden:2022} in combination with BAO and RSD results from galaxy clustering made from the SDSS-II \citep{Howlett:2015}, SDSS-III \citep{Alam-DR11&12:2015}, and SDSS-IV \citep{Bautista:2020,Gil-Marin:2020}. These measurements are shown in Table~\ref{table:prior_results}. Information from the void-galaxy clustering is labelled as voids while information from galaxy-galaxy clustering is labelled as galaxies.

\begin{figure}
\centering
\includegraphics[width=0.44\textwidth]{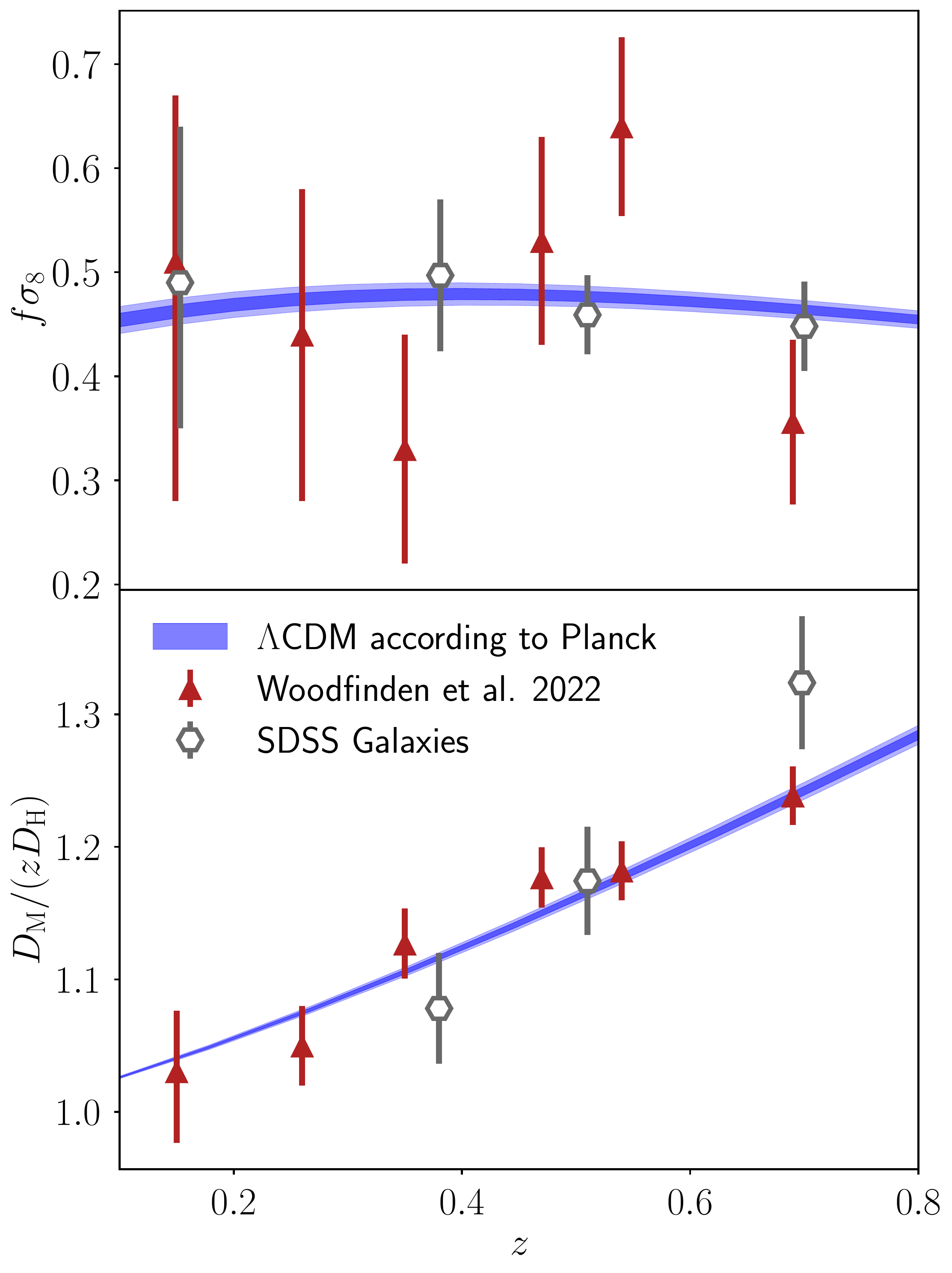}
\caption{Top: Measurements of $f\sigma_8$. Bottom: Measurements of $D_\mathrm{M}/D_\mathrm{H}$, divided by the redshift $z$. Voids in this work are shown as red triangles with associated error bars. Open grey points show the corresponding results from \citet{Howlett:2015,Alam-DR11&12:2015,Bautista:2020} and \citet{Gil-Marin:2020} obtained using BAO measured in the same galaxy samples where applicable (transverse and perpendicular BAO were not separately constrained for MGS at $z=0.15$). The blue shaded band is the $68\%$ and $95\%$ C.L. region obtained from extrapolating the Planck CMB constraints to low redshifts assuming $\Lambda$CDM.}
\label{fig:DMDH-LCDM}
\end{figure}

Figure~\ref{fig:DMDH-LCDM} shows the measurements from both voids and galaxies on $f\sigma_8$ and $D_\mathrm{M}/D_\mathrm{H}$ on SDSS LRG. The best-fit model for $\Lambda$CDM measured by \citet{Planck:2020} is shown for comparison. 

\begin{figure}
\centering
\includegraphics[width=.45\textwidth]{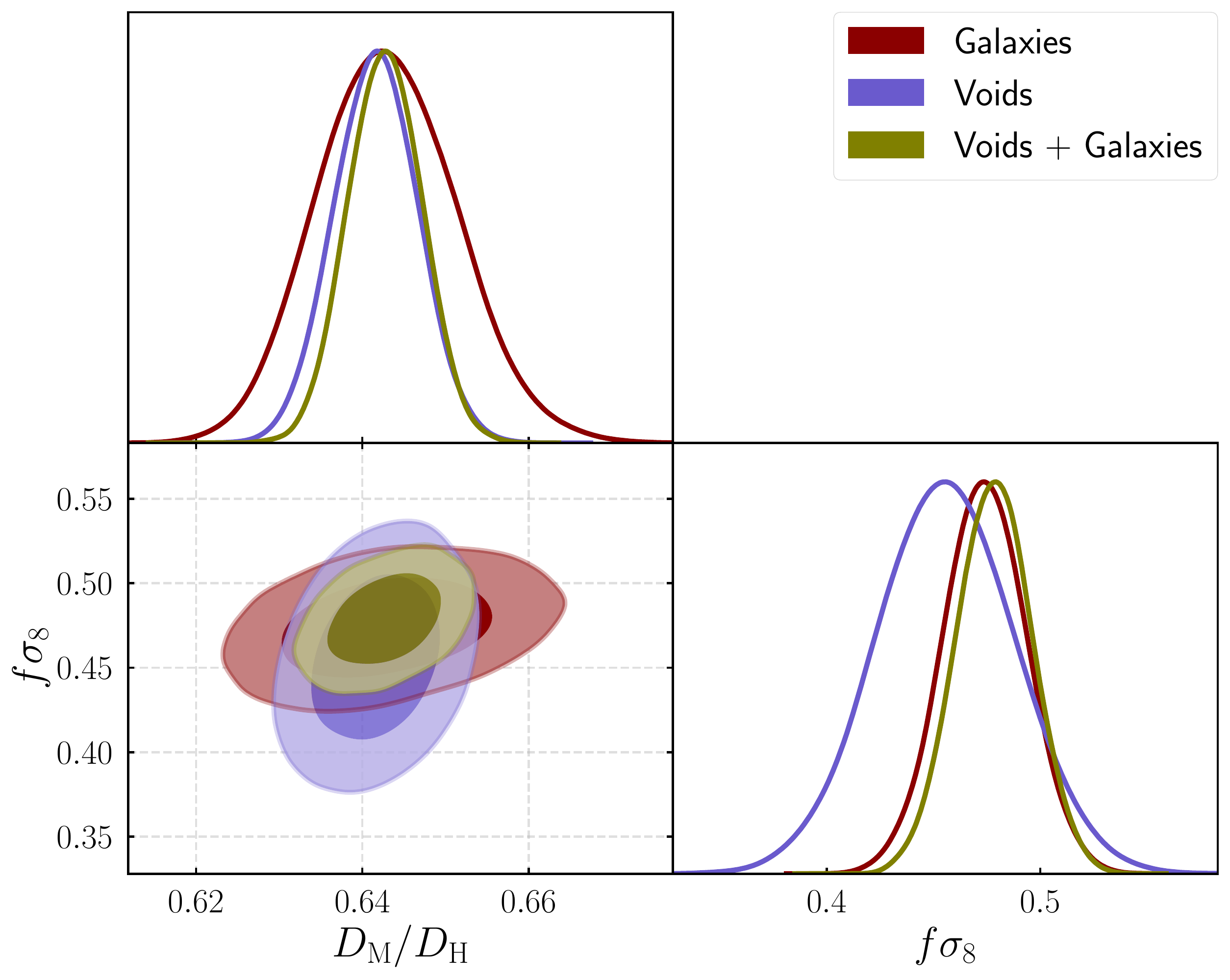}
\caption{
Marginalized compressed constraints on $f\sigma_8$ and $D_\mathrm{M}/D_\mathrm{H}$ at a single redshift $z_\mathrm{eff}=0.52$ (the mean redshift of voids used in this work) obtained from the combination of data shown in Table~\ref{table:prior_results} under the assumption of a flat $\Lambda$CDM cosmological model. Shaded contours show the 68\% and 95\% confidence limit regions and corresponding mean values and marginalized 68\% credible intervals are shown in Table~\ref{table:results}.
}
\label{fig:resultantConstraints}
\end{figure}

Joint fits to both RSD and AP for voids and galaxies combined lead to increased precision in measurements of both the AP effect and $f\sigma_8$ compared to that from voids or galaxies alone. This can be seen in Figure~\ref{fig:resultantConstraints}. The combination of these two measurements, described in more detail in Section~\ref{sec:Combination}, therefore results in a large gain of information than using either of these techniques individually due to the perpendicularity of the likelihood contours. To obtain this combination we have taken information available across a wide range in redshift (see Table~\ref{table:prior_results}) and compressed this, under the assumption of a flat $\Lambda$CDM cosmology, into a single redshift at $z=0.52$ (the mean redshift of voids in this work).

\begin{table*}
\caption{Measurements used in this work. Voids measurements from SDSS samples are from \citet{Woodfinden:2022} and \citet{Nadathur:2020b}, galaxy measurements from SDSS are from \citet{Howlett:2015,Ross:2015,Alam-DR11&12:2015,Bautista:2020} and \citet{Gil-Marin:2020}. The method column refers to the measurement method used: `voids' refers to results from the void-galaxy cross-correlation, and `galaxies' refers to BAO+RSD results from galaxy clustering. $z_\mathrm{eff}$ is the effective redshift at which the measurement was taken. 
Note that galaxy clustering can provide separate measurements of $D_\mathrm{M}$ and $D_\mathrm{H}$ using the BAO scale as a standard ruler while the void-galaxy correlation provides a measurement of their ratio $D_\mathrm{M}/D_\mathrm{H}$ by using a standard shape (i.e. a stack of voids). $D_\mathrm{M}/r_\mathrm{d}$ and $D_\mathrm{H}/r_\mathrm{d}$ information from galaxies has been converted into $D_\mathrm{M}/D_\mathrm{H}$ and $D_\mathrm{V}/r_\mathrm{d}$ using the published correlation coefficients between them.}
\label{table:prior_results}
\begin{tabular}{llllllll}
\hline
 Method   & Redshift Range  & $z_\mathrm{eff}$  & $D_\mathrm{M}/D_\mathrm{H}$ & $D_\mathrm{V}/r_\mathrm{d}$ & $f\sigma_8$ \\ \hline \vspace{-0.28cm}\\
 Voids    & $0.07 < z < 0.2$ & 0.15       & $0.156_{-0.008}^{+0.007}$ & ----------- & $0.51^{+0.16}_{-0.23}$            \\
 Voids    & $0.2 < z < 0.3$ & 0.26      & $0.273 \pm 0.008$ & ----------- & $0.44^{+0.14}_{-0.16}$            \\
 Voids    & $0.3 < z < 0.4$ & 0.35       & $0.397 \pm 0.009$ & ----------- & $0.33 \pm 0.11$            \\
 Voids    & $0.4 < z < 0.5$   & 0.47       & $0.556 \pm 0.011$ & ----------- & $0.53 \pm 0.1$            \\
 Voids    & $0.5 < z < 0.6$ & 0.54      & $0.642 \pm 0.012$ & ----------- & $0.64 \pm 0.077$            \\
 Voids    & $0.6 < z < 1.0$ & 0.69      & $0.868 \pm 0.017$ & ----------- & $0.356 \pm 0.079$            \\
 Galaxies & $0.07 < z < 0.2$ & 0.15      & ----------- & $4.51 \pm 0.14$ & $0.53 \pm 0.16$ \\
 Galaxies & $0.2 < z < 0.5$ & 0.38       & $0.410 \pm 0.016$ & $9.98 \pm 0.11$ & $0.497 \pm 0.045$ \\
 Galaxies & $0.4 < z < 0.6$ & 0.51       & $0.599 \pm 0.021$ & $12.67 \pm 0.13$ &$0.459 \pm 0.038$ \\
 Galaxies & $0.6 < z < 1.0$ & 0.70       & $0.924 \pm 0.035$ & $16.26 \pm 0.21$ & $0.473 \pm 0.041$ \\ \hline
\end{tabular}
\end{table*}

\subsection{MGS}\label{sec:MGS}

The Main Galaxy Sample (MGS; \citealt{Howlett:2015,Ross:2015}) large-scale structure catalogue contains 63,163 galaxies from SDSS Data Release 7 (DR7 \citealt{Abazajian:2009}) in the redshift range $0.07 < z < 0.2$. Previous work from \citet{Woodfinden:2022} analyzed voids using the watershed-based void-finder \texttt{ZOBOV} \citep{Neyrinck:2008}. $517$ total voids at an effective redshift of $z=0.15$ were found with a median size of $40\ \mathrm{h}^{-1} \mathrm{Mpc}$. Void catalogues were cut at the median size resulting in 258 voids being used in the analysis. As summarized in Table~\ref{table:prior_results}, this work measured $f\sigma_8 = 0.51^{+0.16}_{-0.23}$ and $D_\mathrm{M}/D_\mathrm{H} = 0.156^{+0.007}_{-0.008}$. \citet{Howlett:2015} analysed galaxies in this sample including measurements of RSD from the two-point correlation function as well as Alcock-Paczynsi measurements from BAO. This paper did not split Alcock-Paczynski measurements into $D_\mathrm{M}/r_\mathrm{d}$ and $D_\mathrm{H}/r_\mathrm{d}$ as was done for BOSS and eBOSS analysis. Instead, measurements are available for the volume-averaged distance $D_\mathrm{V} = (z D_\mathrm{M}^2 D_\mathrm{H})^{1/3}$ finding $f\sigma_8 = 0.53 \pm 0.16$ and $D_\mathrm{V}/r_\mathrm{d} = 4.51 \pm 0.14$. 

Along with the MGS data, we make use of mock catalogues \citep{Howlett:2015} produced to match the footprint, redshift distribution, and clustering of the MGS data to create our cross-covariance matrix described in Section~\ref{sec:Combination}. 1000 mock catalogues were created out of 500 independent dark matter simulations at $z=0.15$ using the \texttt{PICOLA} algorithm \citep{Howlett:2015b}. Fiducial cosmological parameters $\Omega_\mathrm{m} = 0.31$, $\Omega_\mathrm{b} = 0.048$, $h=0.67$, $\sigma_8 = 0.83$, and $n_\mathrm{s} = 0.96$ were used. Mock galaxies were assigned to halos using halo abundance matching and parameters were chosen to match the galaxy clustering of the MGS data. The survey mask and selection function were also matched to MGS galaxies. Previous work in \citet{Woodfinden:2022} analyzed 1000 of these to test for systematic errors. 

\subsection{BOSS} \label{sec:boss}

The Baryon Oscillation Spectroscopic Survey (BOSS \citealt{Dawson:2013}) of SDSS-III \citep{Eisenstein:2011} surveyed over 1.5 million different objects. Large-scale structure catalogues were released in Data Release 12 (DR12 \citealt{Alam-DR11&12:2015}) and surveyed a redshift extent of $0.2 \lesssim z \lesssim 0.75$. Data was released in two catalogues (LOWZ and CMASS) with varying (overlapping) sky footprints; the results used in this work combine these two samples before splitting into redshift shells. \citet{Woodfinden:2022} analyzed voids in the redshift range $0.2 < z < 0.6$. This work ran a watershed-based void-finder \texttt{ZOBOV} on BOSS data before splitting the voids into 4 redshift shells $0.2 < z < 0.3$, $0.3 < z < 0.4$, $0.4 < z < 0.5$, and $0.5 < z < 0.6$. A total of 8961 voids were found with a median void size of approximately $48\ \mathrm{h}^{-1} \mathrm{Mpc}$. A void size cut of the median void size was imposed resulting in 4480 voids used in the analysis. Measurements on $f\sigma_8$ and $D_\mathrm{M}/D_\mathrm{H}$ are summarized in Table~\ref{table:prior_results}. \citet{Alam:2017} analysed BOSS galaxies measuring $D_\mathrm{M}/r_\mathrm{d}$ and $D_\mathrm{H}/r_\mathrm{d}$ from the BAO method and $f\sigma_8$ from RSD. Measurements are analysed in 3 partially overlapping redshift shells, $0.2 < z < 0.5$, $0.4 < z < 0.6$, and $0.5 < z < 0.75$. The highest redshift shell overlaps entirely with a combination of the $0.4 < z < 0.6$ and $0.5 < z < 0.75$ shells in the redshift range $0.6 < z < 1.0$ (discussed in Section~\ref{sec:eboss}) and so is not included in this work. Measurements are summarized in Table~\ref{table:prior_results}. $D_\mathrm{M}/r_\mathrm{d}$ and $D_\mathrm{H}/r_\mathrm{d}$ information has been converted into $D_\mathrm{M}/D_\mathrm{H}$ and $D_\mathrm{V}/r_\mathrm{d}$ using the published correlation coefficients between them.

Along with the BOSS data, we make use of the Patchy mocks to create our cross-covariance matrix described in Section~\ref{sec:Combination}. The Patchy mocks are a set of 1000 independent mock catalogues created on the lightcone using the fast approximate \texttt{Patchy} algorithm \citep{Kitaura:2014}. The Patchy mocks were created to match the clustering and survey properties of BOSS galaxies \citep{Kitaura:2016} with fiducial cosmological parameters $\Omega_\mathrm{m} = 0.31$, $\Omega_\mathrm{b} = 0.0482$, $h=0.6777$, $\sigma_8 = 0.8225$, and $n_\mathrm{s} = 0.96$. Mock galaxies were assigned to halos using halo abundance matching and parameters were chosen to match the galaxy cluster properties measured through the monopole and quadrupole moments. The survey mask and selection function were also matched to BOSS galaxies. Previous work in \citet{Woodfinden:2022} analyzed 250 of these to test for systematic errors. We use this same 250 mock sub-sample to construct our cross-covariance matrix. 

\subsection{eBOSS} \label{sec:eboss}

The extended Baryon Oscillation Spectroscopic Survey (eBOSS \citealt{Dawson:2016}) of SDSS-IV \citep{Blanton:2017} surveyed over 375,000 different objects over two hemispheres. Large-scale structure catalogues were released in Data Release 16 (DR16, \citealt{Ahumada:2020}) and surveyed a redshift extent of $0.6 \lesssim z \lesssim 1.0$. Analysis of this sample used in this work combined measurements from eBOSS with galaxies from BOSS DR12 that overlap this redshift range. \citet{Nadathur:2020b} analysed voids in this sample using watershed-based void-finder \texttt{ZOBOV} finding a total of 4706 voids with a median void size of $49\ \mathrm{h}^{-1} \mathrm{Mpc}$. A median size cut was applied resulting in 2341 voids used in the analysis. This work measured $f\sigma_8 = 0.356 \pm 0.079$ and $D_\mathrm{M}/D_\mathrm{H} = 0.868 \pm 0.017$. Galaxies in this sample were analysed by \citet{Bautista:2020} and \citet{Gil-Marin:2020} using BAO and RSD features from the galaxy two-point correlation function and power spectrum to infer geometrical and dynamical cosmological constraints of $f\sigma_8 = 0.473 \pm 0.041$, $D_\mathrm{M}/r_\mathrm{d} = 17.65\pm0.3$, and $D_\mathrm{H}/r_\mathrm{d} = 19.78 \pm 0.46$. 

Along with eBOSS data, we make use of the EZmocks to create our cross-covariance matrix described in Section~\ref{sec:Combination}. The EZmocks are a set of 1000 independent mock catalogues created on the lightcone using the \texttt{EZmock} algorithm \citep{Chuang:2015}. The EZmocks were created to match the galaxy clustering and survey properties of eBOSS galaxies with fiducial parameters $\Omega_\mathrm{m} = 0.307$, $\Omega_\mathrm{b} = 0.0482$, $h=0.6777$, $\sigma_8 = 0.8225$, and $n_\mathrm{s} = 0.96$. The \texttt{EZmock} algorithm is a fast-approximate Zeldovich method (similar to the \texttt{Patchy} algorithm) along with deterministic and stochastic biased relations, a probability density function mapping scheme and addition correction to account for RSD \citep{Zhao:2021}.  

\section{Combination of Results and Likelihoods}
\label{sec:Combination}

\begin{figure*}
    \centering
    \includegraphics[width=0.85\textwidth]{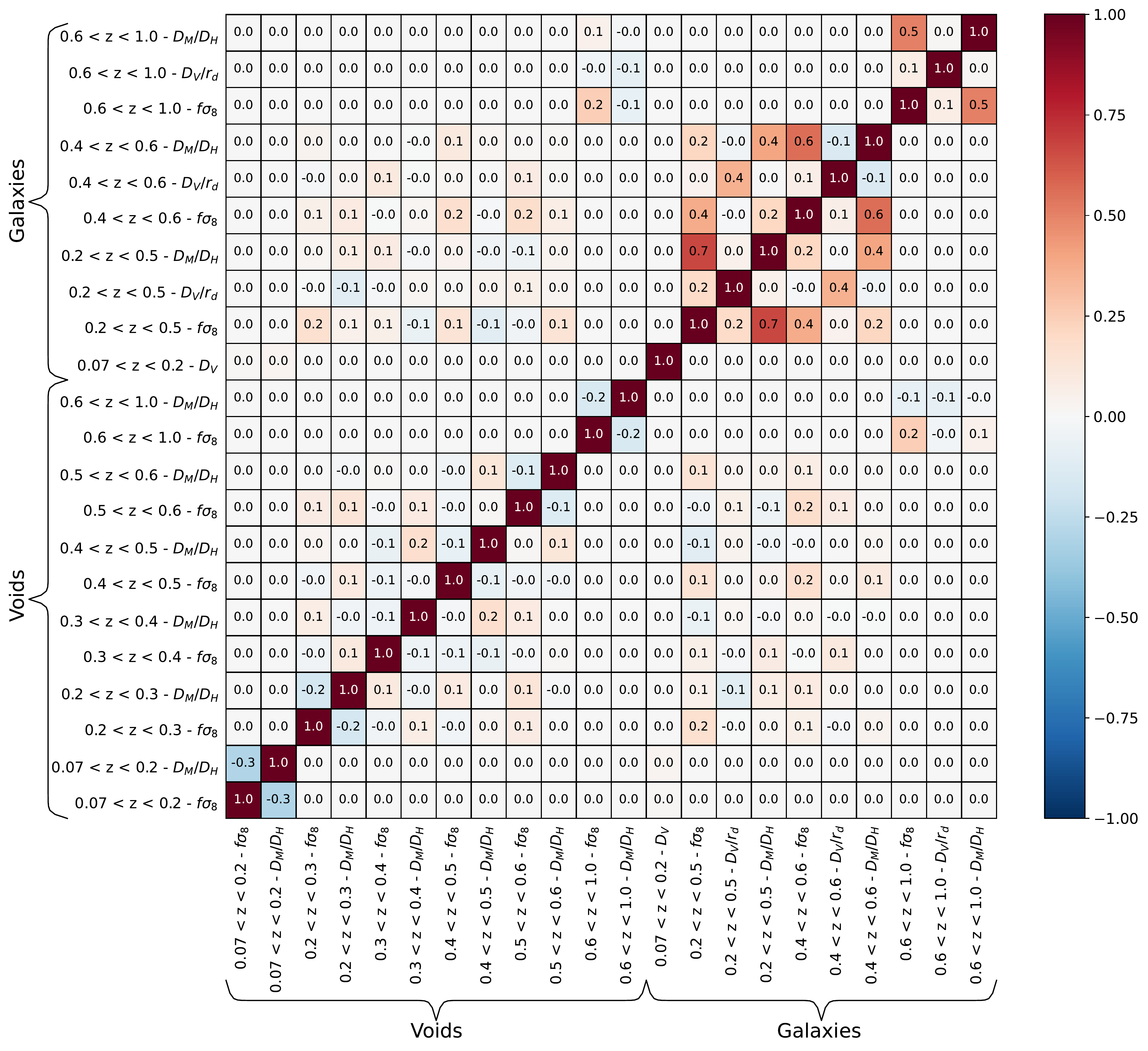}
    \caption{Correlation coefficients between measured values of $f\sigma_8$, $D_\mathrm{M}/D_\mathrm{H}$, and $D_\mathrm{V}$ obtained from voids and galaxies in various redshift bins from SDSS LRGs. The cross correlation between measurements taken from different mock catalogues is set to 0 to reduce noise in unrelated measurements.}
    \label{fig:CrossCorrelation} 
\end{figure*}

We combine void-galaxy and galaxy-galaxy clustering measurements to take advantage of the significant gain of information from the complementary directions of parameter contours. As both methods are applied to the same data, the large-scale structure could conceivably affect both measurements. We account for any covariance between the two using the method from \citet{Alam:2017} and \citet{Sanchez:2017} and previously applied to the void-galaxy correlation function in combination with galaxy clustering in \citet{Nadathur:2019a} and \citet{Nadathur:2020b}. A cross-covariance matrix, $\Psi$ is constructed for all observables using measurements made on mock catalogues described in Section~\ref{sec:data}. The covariance between measurements made on different mock catalogues (ex. between MGS mocks in the redshift range $0.07 < z < 0.2$ and eBOSS mocks in the redshift range $0.6 < z < 1.0$) will only be non-zero due to random noise and are set equal to 0. This is typical of combined analyses and the correlations for non-overlapping galaxy samples (this is \emph{not} the case for the BOSS galaxy samples in this work) are close to zero \citep{Sanchez:2016}. Figure~\ref{fig:CrossCorrelation} shows that the cross-correlation between different void redshift slices is small. It is not possible to measure the cross-correlation directly between SDSS samples as this would require mock catalogues with a combination of volume and resolution which is not currently possible for our samples.

We combine all observables into one data vector $\vec{D}$, and at each point in theory space being sampled we calculate the $\chi^2$ value as
\begin{equation}
    \chi^2 = (\vec{D} - \vec{T})^t \Psi (\vec{D} - \vec{T})
\end{equation}
where $\vec{T}$ is a vector containing the value of all observables at a particular point in the theory space being sampled.

The data vectors used have already included a prior designed to match the Bayesian results to frequentist expectations to first order allowing for errors in the covariance matrix used when fitting to the correlation functions. From these fits, we only use the covariant parameter measurements: hence only needing a smaller number of mocks, which are only used to estimate the covariance between measurements in the same redshift bins. As such we can assume these measurements as a gaussian fit and so use an appropriate likelihood function \citep{Woodfinden:2022,Percival:2021}. To explore the model parameter space we use MCMC sampling implemented using the \texttt{Cobaya} sampling package \citep{Torrado:2019,Torrado:2021} and use a mixture of \texttt{CAMB}\footnote{https://camb.info/} and \texttt{CLASS}\footnote{https://lesgourg.github.io/class\_public/class.html} as the underlying cosmological codes \citep{Lewis:2000,Lesgourgues:2011}. 

Figure~\ref{fig:CrossCorrelation} shows the cross-correlation matrix corresponding to the cross-covariance matrix. Only a weak correlation can be seen between measurements of $f\sigma_8$ and $D_\mathrm{M}/D_\mathrm{H}$ as measured by voids. While results are reported for galaxy clustering in the parameters $f\sigma_8$, $D_\mathrm{M}$, and $D_\mathrm{H}$ we perform a change of basis to $f\sigma_8$, $D_\mathrm{M}/D_\mathrm{H}$, and $D_\mathrm{V}$ using the published correlation coefficients between them.

Results from BAO+RSD for galaxies show a strong correlation in measured parameters. We see that for all redshift bins considered in this work, there is a strong positive correlation between $f\sigma_8$ and $D_\mathrm{M}/D_\mathrm{H}$, and little correlation elsewhere. We also see similar trends when looking at the correlations between redshift bins $0.2 < z < 0.5$ and $0.4 < z < 0.6$. This significant correlation between these bins is expected as these come from the same mock catalogues which are later split into redshift shells as well as overlap in the $0.4 < z < 0.5$ range. 

No significant correlation can be seen between results from voids and galaxies when looking at data measured from the same mock catalogues. A combination of measurements from voids and galaxies will therefore result in a significant gain in information compared to each measurement individually. 

The results from combining information from void-galaxy and galaxy-galaxy correlations on measured parameters $f\sigma_8$, $D_\mathrm{M}$, $D_\mathrm{H}$, and $D_\mathrm{M}/D_\mathrm{H}$ are shown in Table~\ref{table:resultantConstraints}. These are found assuming a $\Lambda$CDM cosmological model. The effective redshift of voids in the sample is calculated as the weighted sum
\begin{equation}
    z_\mathrm{eff} = \dfrac{\Sigma_\mathrm{ij} \left( \frac{Z_\mathrm{i} + z_\mathrm{j}}{2} w_\mathrm{j} \right)}{\Sigma_\mathrm{ij} w_\mathrm{j}},
\end{equation}
where $Z_i$ is the redshift of the void centre, $z_j$ is the galaxy redshift, $w_j$ is the associated galaxy systematic weight, and the sum extends over all void-galaxy pairs up to the maximum separation considered, $120 h^{-1}$Mpc. 

Parameter contours plotted in Figure~\ref{fig:resultantConstraints} demonstrate the significant gain of information from the combination of these results; this gain of information is due to the orthogonality of degeneracy directions between void-galaxy and galaxy-galaxy correlations as well as the weak correlation between these two methods. 

\begin{table*}
\caption{Final results for the combination of void-galaxy and galaxy-galaxy clustering. We report results at the effective redshift of voids $z_\mathrm{eff}$  The combination of measurements shown is made by assuming a $\Lambda$CDM cosmological model. Galaxies refer to constraints found through measurement of the BAO+RSD and void refers to constraints found through void-galaxy cross-correlation. Reported errors are the 68\% credible intervals.}
\label{table:resultantConstraints}
\begin{tabular}{lllllll}
\hline
Tracer   & Redshift range    &  $z_\mathrm{eff}$    & $f\sigma_8$ & $D_\mathrm{M}/r_\mathrm{d}$ & $D_\mathrm{H}/r_\mathrm{d}$ & $D_\mathrm{M}/D_\mathrm{H}$ \\ \hline
Voids + Galaxies & $0.07 < z < 1.0$ & $0.52$  & $0.464\pm 0.017$ & $17.36\pm 0.13$ & $19.70\pm 0.21$ & $0.8655\pm 0.0071$ \\
Voids + Galaxies & $0.4 < z < 0.6$ & $0.51$  & $0.479\pm 0.028$ & $13.40\pm 0.15$ & $22.36\pm 0.30$ & $0.6410\pm 0.0075$ \\
Voids + Galaxies &   $0.2 < z < 0.5$ & $0.39$ & $0.513\pm 0.028$ & $10.35\pm 0.13$ & $24.18\pm 0.34$ & $0.4282\pm 0.0051$\\ \hline
\end{tabular}
\end{table*}

\section{Results}\label{sec:Results}

We split the discussion of our results by the cosmological model assumed when running MCMC fits to the data.  Relevant cosmological parameters are shown in each section. 

\subsection{$\Lambda$CDM}

For a base $\Lambda$CDM model the AP information constrains $\Omega_\mathrm{m}$, as shown in Table~\ref{table:results}. If we do not know $r_\mathrm{d}$, then the BAO measurements reduce to only the relative constraints from the AP effect, while without an independent constraint on $\sigma_8$, the RSD measurement also becomes a relative measure, and only depends on the matter density. 

Of note in Table~\ref{table:results}, the uncertainty on $\Omega_\mathrm{m}$ from voids+galaxies is 43\% lower than that from galaxies alone due to the increase in precision in measurements of $D_\mathrm{M}/D_\mathrm{H}$. The measured value of $\Omega_\mathrm{m}$ from the full SDSS results is consistent with results from \citet{Planck:2020} within $1.6\sigma$, however, this is in increased tension as that from galaxies alone agrees with Planck within $1 \sigma$, due to a larger uncertainty. 

The combination of Planck and SN measures $\Omega_m = 0.3143\pm 0.0079$ and $H_0 = 67.43\pm 0.57$. The additional information from SDSS (voids+galaxies) results in a $30\%$ reduction in the measured error on $\Omega_\mathrm{m}$ and a $28\%$ reduction in the measured error on $H_0$.

\begin{table*}
\caption{Final parameter constraints from a combination of different probes. Voids refer to measurements made through the void-galaxy correlation function. Galaxies refer to constraints found through measurement of the BAO peak position and peculiar velocities of galaxies. Voids and Galaxies are both measurements from SDSS. CMB refers to measurements of the cosmic microwave background from \citet{Planck:2020}. SN refers to measurements from the Pantheon SN Ia sample \citep{Scolnic:2017}. Reported errors are the 68\% credible intervals. Voids provide constraints in a narrow band in the $\Omega_\mathrm{m} - \Omega_\Lambda$ plane in open cosmologies, shown in Figure~\ref{fig:oCDM} however are not able to constrain $\Omega_\mathrm{m}$ or $\Omega_\Lambda$ and are not shown. Void, Galaxies, and Voids + Galaxies without additional CMB +SN data were unable to provide constraints on $w(z)CDM$ and $ow(z)CDM$ models within a reasonable prior volume and so are excluded from this table.}
\label{table:results}
\begin{tblr}{lllllll}
\hline
Measurement Techniques & $\Omega_\mathrm{m}$ & $\Omega_\Lambda$ & $\Omega_\mathrm{k}$ & $w_0$ & $w_\mathrm{a}$ & $H_0$\\ \hline
$\mathbf{\Lambda CDM}$ & & & & & & \\ 
Voids   & $0.354\pm 0.029$ & ----------- & ----------- & ----------- & ----------- & -----------         \\
Galaxies  & $0.360^{+0.038}_{-0.045}$ & ----------- & ----------- & ----------- & -----------  & -----------        \\
Voids + Galaxies   & $0.356\pm 0.024$ & ----------- & ----------- & ----------- & -----------  & -----------        \\
Voids + CMB + SN & $0.3170\pm 0.0069$ & ----------- & ----------- & ----------- & -----------    & $67.24\pm 0.50$    \\ 
Galaxies + CMB + SN &  $0.3117\pm 0.0056$ & ----------- & ----------- & ----------- & -----------    & $67.62\pm 0.42$      \\
Voids + Galaxies + CMB + SN & $0.3127\pm 0.0055$ & ----------- & ----------- & ----------- & ----------- & $67.55\pm 0.41$       \\ \hline[dashed]
$\mathbf{w CDM}$ & & & & & & \\ 
Voids   & $0.330^{+0.11}_{-0.033}$ & ----------- & ----------- & $-1.17^{+0.64}_{-0.31}$ & -----------  & -----------        \\
Galaxies  & $0.359\pm 0.044$ & ----------- & ----------- & $-1.91^{+0.45}_{-0.85}$ & -----------   & -----------       \\
Voids + Galaxies   & $0.391^{+0.028}_{-0.021}$ & ----------- & ----------- & $-1.50^{+0.43}_{-0.28}$  & -----------     & -----------     \\
Voids + CMB + SN & $0.3148\pm 0.00715$ & ----------- & ----------- & $-0.906^{+0.10}_{-0.071}$ & -----------     & $67.39\pm 0.52$     \\ 
Galaxies + CMB + SN &  $0.3149\pm 0.0066$ & ----------- & ----------- & $-0.960\pm 0.045$ & -----------   & $67.39\pm 0.49$      \\
Voids + Galaxies + CMB + SN & $0.3172\pm 0.0061$ & ----------- & ----------- & $-0.930\pm 0.039$ & -----------   & $67.22\pm 0.45$      \\ \hline[dashed]
$\mathbf{w(z) CDM}$ & & & & & & \\ 
Voids + CMB + SN & $0.3152\pm 0.0072$ & ----------- & ----------- & $-1.02\pm 0.34$ & $>-0.54$ & $67.37\pm 0.52$\\ 
Galaxies + CMB + SN &  $0.3139\pm 0.0070$ & ----------- & ----------- & $-0.998^{+0.097}_{-0.11}$ & $0.16^{+0.51}_{-0.36}$ & $67.46\pm 0.51$ \\
Voids + Galaxies + CMB + SN & $0.3152\pm 0.0067$ & ----------- & ----------- & $-0.995\pm 0.094$ & $0.27^{+0.40}_{-0.33}$    & $67.37\pm 0.49$     \\ \hline[dashed]
$\mathbf{o CDM}$ & & & & & &\\ 
Voids   & ---------- & ---------- & ---------- & ----------- & -----------   & -----------      \\
Galaxies  & $0.358^{+0.045}_{-0.058}$ & $0.641^{+0.083}_{-0.069}$ & $0.001^{+0.094}_{-0.11}$ & ----------- & -----------   & -----------       \\
Voids + Galaxies  & $0.354^{+0.045}_{-0.055}$ & $0.638\pm 0.049$ & $0.008^{+0.098}_{-0.083}$ & ----------- & -----------   & -----------       \\
Voids + CMB + SN & $0.335\pm 0.015$ & $0.671\pm 0.011$ & $-0.0061\pm 0.0045$ & ----------- & ----------- & $65.1\pm 1.6$ \\ 
Galaxies + CMB + SN &  $0.3114\pm 0.0063$ & $0.6881\pm 0.0058$ & $0.0004\pm 0.0020$ & ----------- & ----------- & $67.72\pm 0.66$ \\
Voids + Galaxies + CMB + SN & $0.3120\pm 0.0060$ & $0.6873\pm 0.0055$ & $0.0007^{+0.0019}_{-0.0018}$ & ----------- & ----------- & $67.71\pm 0.63$ \\ \hline[dashed]
$\mathbf{o w CDM}$& & & & & & \\ 
Voids   & $0.385^{+0.082}_{-0.13}$ & $0.657^{+0.077}_{-0.14}$ & $-0.04^{+0.22}_{-0.15}$ & $-1.29^{+0.58}_{-0.32}$ & -----------    & -----------      \\
Galaxies  & $0.326^{+0.046}_{-0.064}$ & $0.582^{+0.063}_{-0.077}$ & $0.093^{+0.11}_{-0.083}$ & $-1.98^{+0.46}_{-0.86}$ & -----------  & -----------      \\
Voids + Galaxies   & $0.331^{+0.067}_{-0.094}$ & $0.607^{+0.078}_{-0.12}$ & $0.06^{+0.18}_{-0.13}$ & $-1.41^{+0.70}_{-0.31}$ & -----------   & -----------    \\
Voids + CMB + SN & $0.332\pm 0.017$ & $0.674\pm 0.013$ & $-0.0052\pm 0.0048$ & $-0.954^{+0.14}_{-0.075}$ & -----------  & $65.5^{+1.7}_{-1.9}$     \\ 
Galaxies + CMB + SN &  $0.323\pm 0.011$ & $0.6800\pm 0.0088$ & $-0.0029^{+0.0036}_{-0.0032}$ & $-0.910^{+0.078}_{-0.069}$ & ----------- & $66.4\pm 1.2$      \\
Voids + Galaxies + CMB + SN & $0.3239\pm 0.0085$ & $0.6791\pm 0.0069$ & $-0.0031\pm 0.0028$ & $-0.889\pm 0.052$ & -----------  & $66.29\pm 0.93$        \\ \hline[dashed]
$\mathbf{o w(z) CDM}$ & & & & & &\\ 
Voids + CMB + SN & $0.360^{+0.021}_{-0.026}$ & $0.652^{+0.019}_{-0.016}$ & $-0.0126^{+0.0074}_{-0.0057}$ & $-1.22\pm 0.43$ & $>-0.48$ & $62.8\pm 2.1$  \\
Galaxies + CMB + SN &  $0.322\pm 0.014$ & $0.680\pm 0.011$ & $-0.0027^{+0.0042}_{-0.0038}$ & $-0.91\pm 0.17$ & $-0.02^{+0.56}_{-0.48}$ & $66.5\pm 1.6$ \\
Voids + Galaxies + CMB + SN & $0.323\pm 0.011$ & $0.6795\pm 0.0087$ & $-0.0030\pm 0.0034$ & $-0.90\pm 0.14$ & $0.02^{+0.49}_{-0.42}$& $66.3\pm 1.2$  \\\hline
\end{tblr}
\end{table*}

\subsection{wCDM}

In this section, we consider a $w$CDM model, where we expand the $\Lambda$CDM model to consider a dark energy component with equation of state $w_0 \equiv p/\rho$, where $p$ is the dark energy pressure and $\rho$ is the dark energy density. In the standard $\Lambda$CDM model, $w_0 = -1$, describing dark energy as a cosmological constant. A value of $w_0 < -1/3$ is required for an accelerating expansion of the Universe while a value of $w_0 < -1$ means that dark energy density increases as the Universe expands. 

\begin{figure*}
\centering
\includegraphics[width=.48\textwidth]{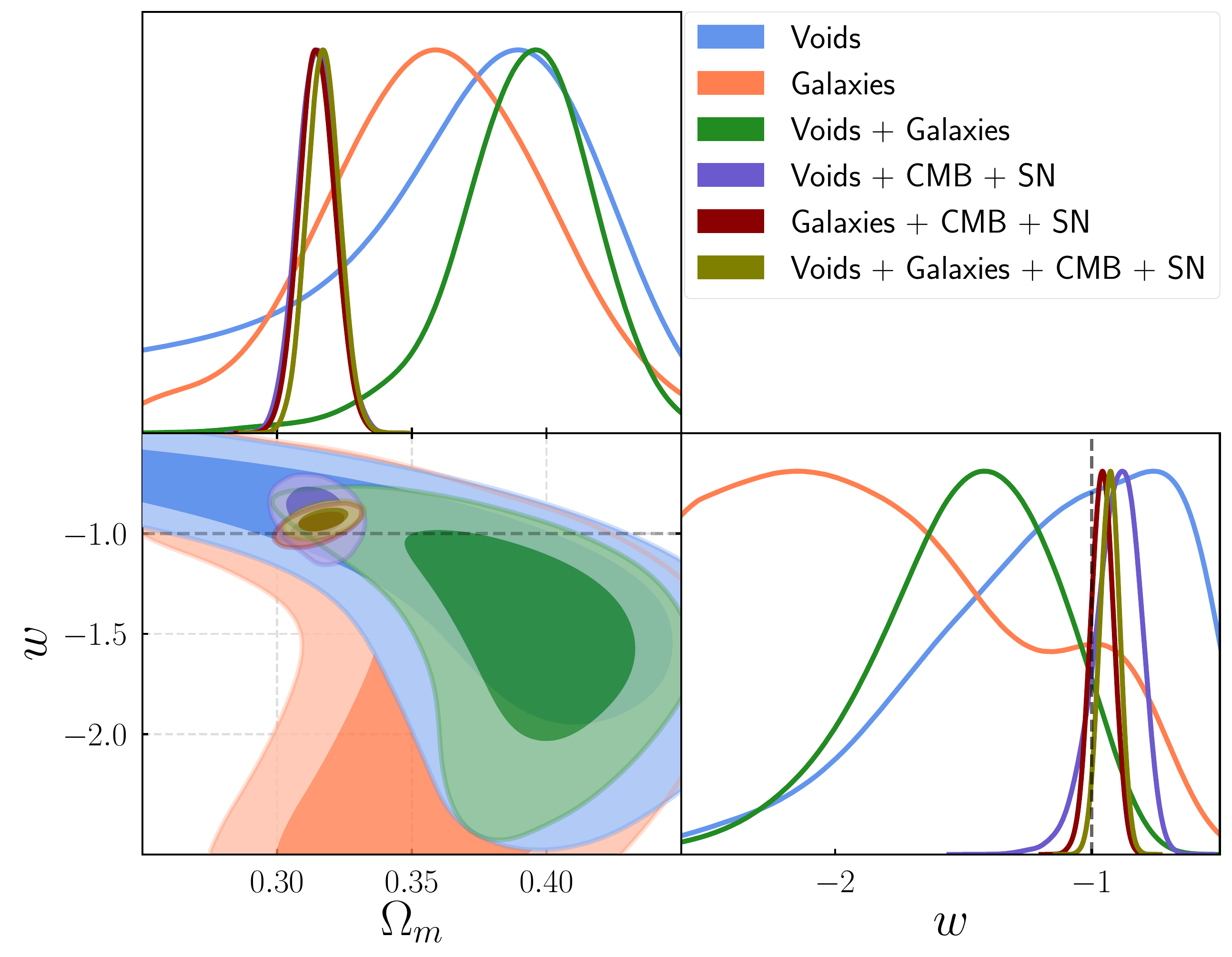}\quad
\includegraphics[width=.48\textwidth]{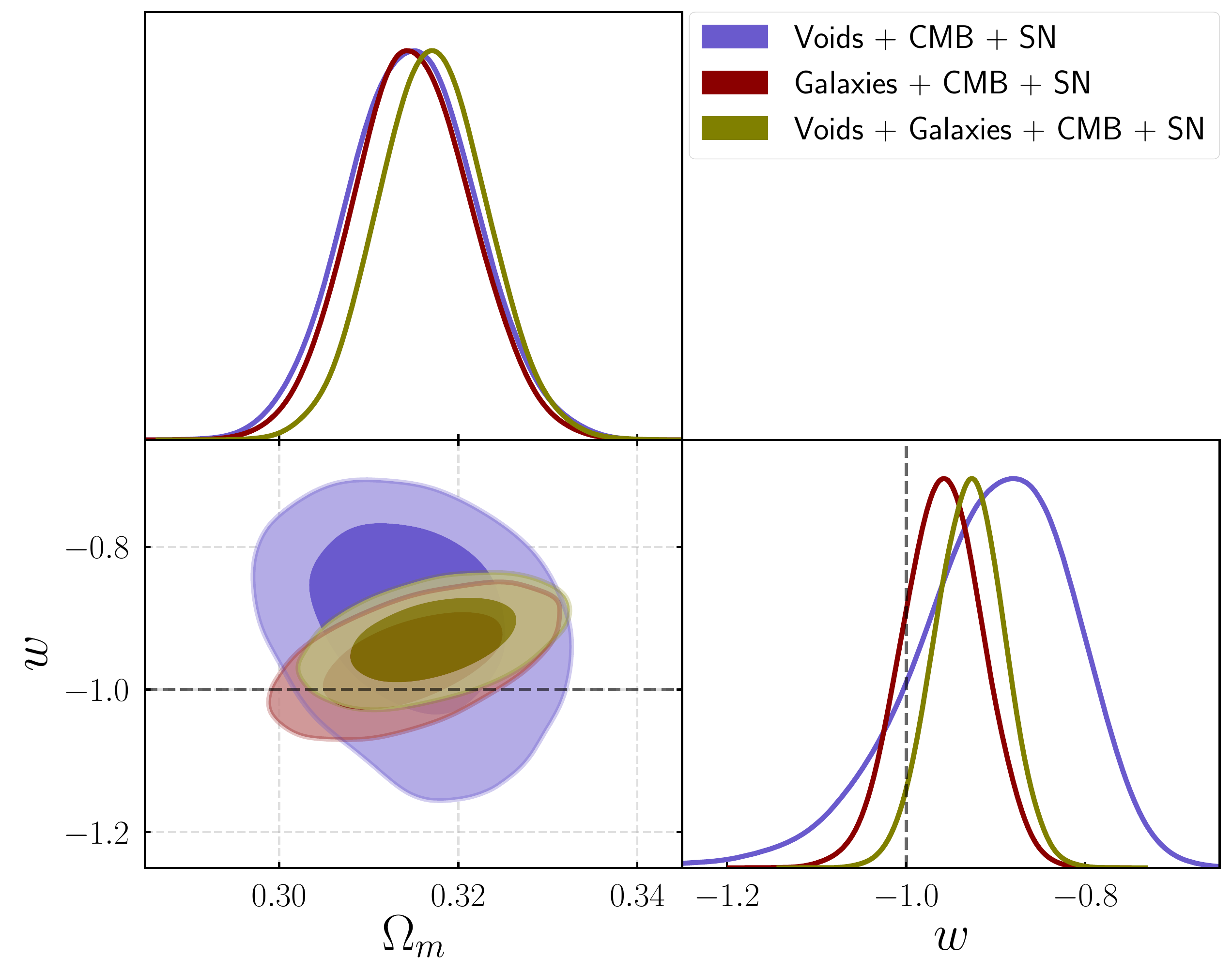}
\caption{Marginalized posterior constraints on cosmological parameters under the assumption of a $w$CDM cosmology. Shaded contours show the 68\% and 95\% confidence limit regions and corresponding mean values and marginalized 68\% credible intervals are shown in Table~\ref{table:resultantConstraints}. On the left, we show results from SDSS Voids and Galaxies alone as well as the combination of these two datasets. We also show these three combinations with the addition of CMB measurements from Planck TT,TE,EE+lowE+lensing and SN measurements from the Pantheon SN Ia sample. The right plot shows only the combination of SDSS data with Planck and Pantheon. The addition of void information results in a significant gain of information both with and without the inclusion of CMB + SN data. The dashed line corresponds to the $\Lambda$CDM prediction of $w=-1$.}
\label{fig:wCDM}
\end{figure*}

An uninformative, flat prior of $-3 < w < -1/3$ is imposed. Figure~\ref{fig:wCDM} shows results for this cosmology and Table~\ref{table:results} lists the mean values and marginalized 68\% credible intervals obtained for various combinations of measurement techniques. Results show reasonable agreement with standard $\Lambda$CDM cosmology where $w=-1$.

\begin{figure*}
\centering
\includegraphics[width=.48\textwidth]{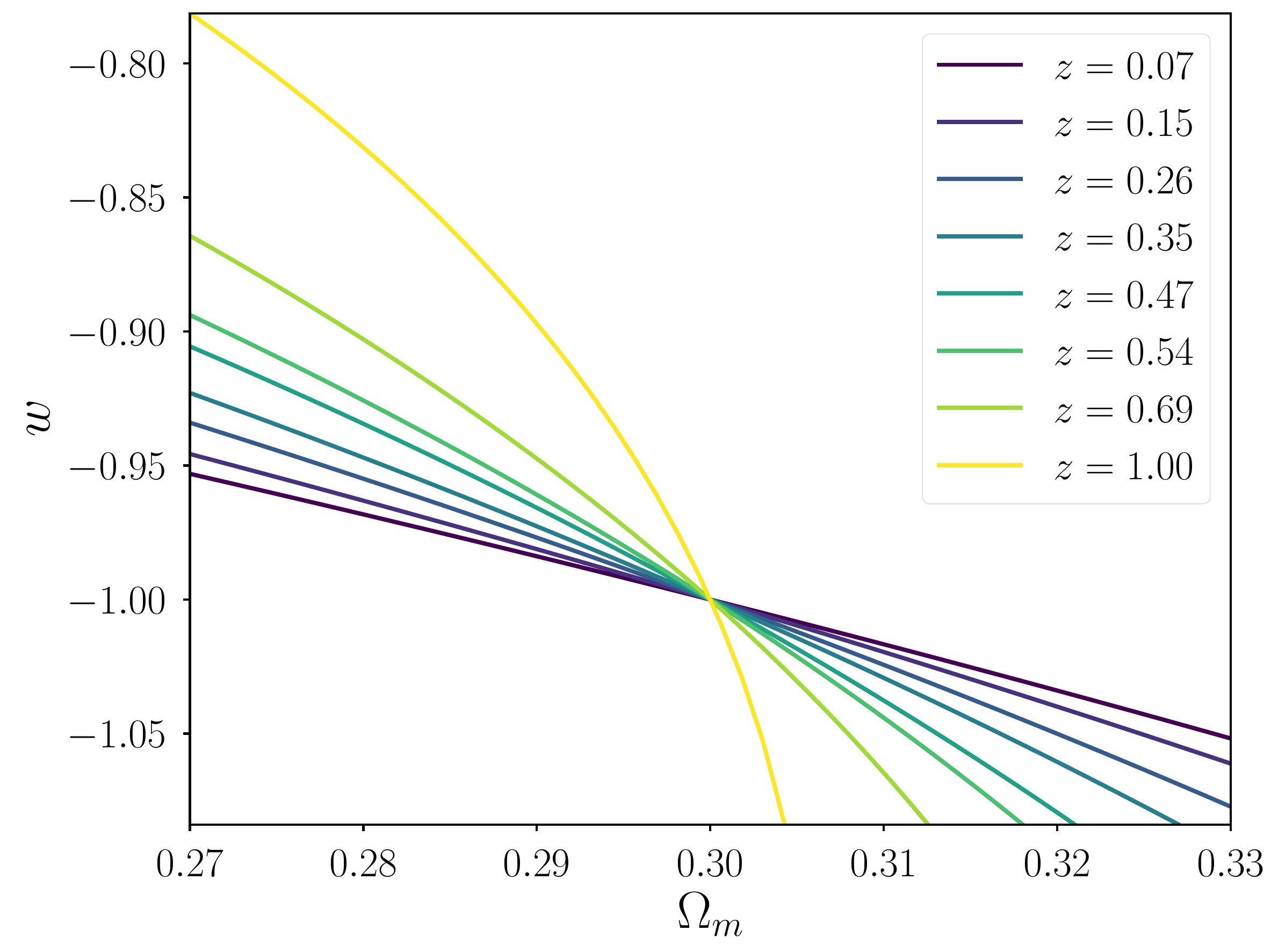}\quad
\includegraphics[width=.48\textwidth]{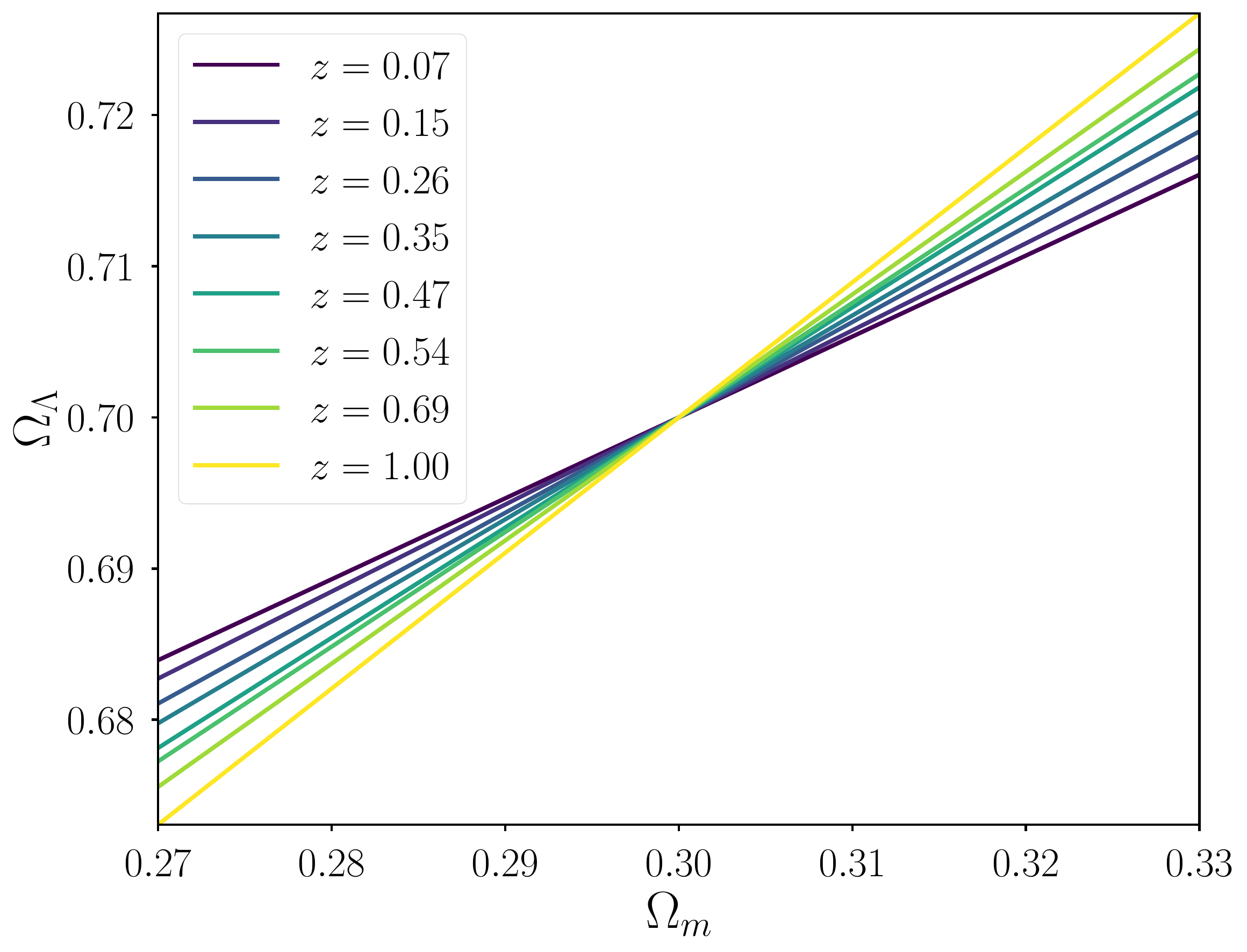}
\caption{The loci of models with $D_\mathrm{M}/D_\mathrm{H}$ matching that of a cosmological model with $\Omega_\mathrm{m} = 0.3$, $\Omega_\Lambda = 0.7$, and $w=-1$, for a wide range in redshift. Left shows loci for a $w$CDM cosmology while right shows loci for an $o$CDM cosmology. Redshifts from $z=0.07$ up to $z=1.0$ are chosen to match the average redshift of the slices described in Table~\ref{table:prior_results} as well as the minimum and maximum redshifts of the data used.}
\label{fig:gradient}
\end{figure*}

For a given cosmology and redshift, the constraints on $D_\mathrm{M}/D_\mathrm{H}$ have a strong degeneracy following the locus of models that predict the same theoretical value of $D_\mathrm{M}/D_\mathrm{H}$. The direction of this locus changes with redshift as shown in Figure~\ref{fig:gradient} for models where $w$ and $\Omega_\mathrm{m}$ are varied, as well as models where $\Omega_\mathrm{m}$ and $\Omega_\Lambda$ are varied. The change in locus depends on Equation~\ref{eq:DM}. The change in locus means that we significantly improve our constraints by having measurements at multiple redshifts compared with a measurement with an equivalent total error at a single redshift. Thus having void measurements at a series of redshifts is an important resource and significantly improves the cosmological constraints presented here. Measurements from upcoming surveys at high redshift will provide additional constraints on $w$CDM and $o$CDM beyond constraints due to the increased volume that these surveys measure.

The combination of Planck and SN only measures $\Omega_m = 0.307\pm 0.011$, $w = -1.036\pm 0.037$ and $H_0 = 68.3\pm 1.1$. The additional information from SDSS (voids+galaxies) results in a $5\%$ reduction in the measured error on $\Omega_\mathrm{m}$, a $5\%$ increase in the measured error on $w$ and a $59\%$ reduction in the measured error on $H_0$.

\subsection{w(z)CDM}  \label{sec:wCDM}

In this section, we consider adding more flexibility to the equations of state of dark energy. If dark energy is a generic dynamical fluid then the equation of state parameter $w$ should be allowed to vary over time. We adopt an equation of state parameter for dark energy with the functional form $w(a) = w_0 + (1-a) w_\mathrm{a}$. Here $w_0$ and $w_\mathrm{a}$ are fitting parameters and $a$ is the scale factor. In a $\Lambda$CDM model $w_0=-1$ and $w_\mathrm{a}=0$. 

\begin{figure}
\centering
\includegraphics[width=.48\textwidth]{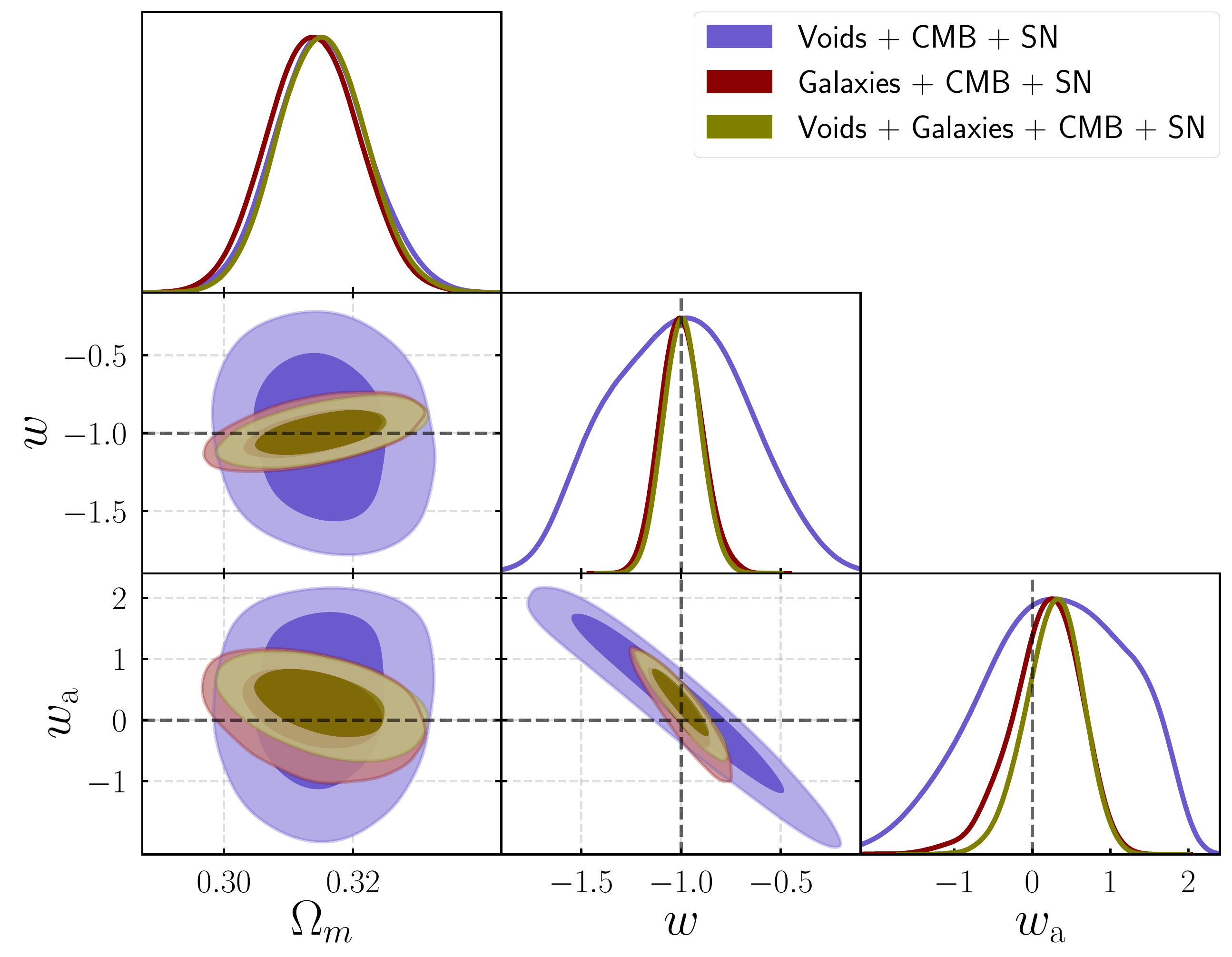}
\caption{Marginalized posterior constraints on cosmological parameters under the assumption of a $ww_\mathrm{a}$CDM cosmology. Shaded contours show the 68\% and 95\% confidence limit regions and corresponding mean values and marginalized 68\% credible intervals are shown in Table~\ref{table:resultantConstraints}. Results are for combinations of SDSS probes (Voids and Galaxies) in combination with CMB measurements from Planck TT,TE,EE+lowE+lensing and SN measurements from the Pantheon SN Ia sample. The Void + CMB + SN constraints can be seen to be hitting the upper boundary of the prior on $w_\mathrm{a}$ so only lower limits are reported. The dashed line corresponds to the $\Lambda$CDM prediction of $w=-1$ and $w_\mathrm{a}=0$.}
\label{fig:wwaCDM}
\end{figure}

Uninformative, flat priors of $-3 < w < -1/3$ and $-3 < w_\mathrm{a} < 2$ are imposed. Figure~\ref{fig:wwaCDM} shows results for this cosmology and Table~\ref{table:results} lists the mean values and marginalized 68\% credible intervals obtained for various combinations of measurement techniques. Only the combination of SDSS data with CMB and SN results from \citet{Planck:2020} and \citet{Scolnic:2017} are shown, as SDSS alone does not have enough constraining power within a reasonable prior volume. All results show agreement with $w_0 = -1$ and $w_\mathrm{a} = 0$, demonstrating consistency with a $\Lambda$CDM model. For the combination of voids with CMB and SN data, the constraints on $w_\mathrm{a}$ can be seen to be hitting the upper bound of the prior; as such only lower limits are given in Table~\ref{table:results}.

\subsection{oCDM}

In this section, we apply an $o$CDM model, where $\Omega_\mathrm{k} \equiv \Omega_\mathrm{m} + \Omega_\Lambda - 1$ is left as a free parameter. In $\Lambda$CDM spatial curvature is considered to be flat (i.e. $\Omega_\mathrm{k} = 0$). A value of $\Omega_\mathrm{k} < 0$ indicates a closed Universe while $\Omega_\mathrm{k} > 0$ indicated an open Universe. 

\begin{figure*}
\centering
\includegraphics[width=.48\textwidth]{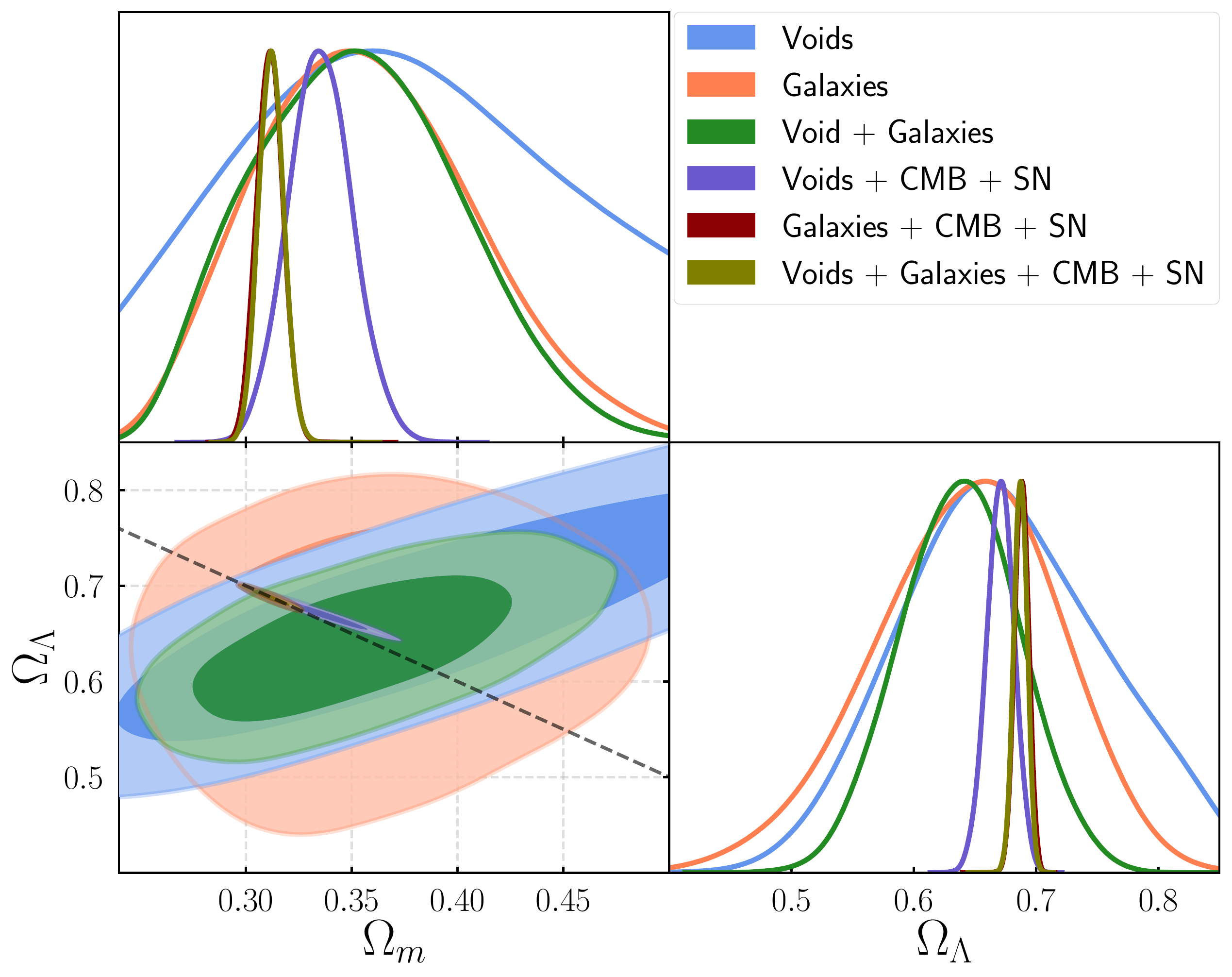}\quad
\includegraphics[width=.48\textwidth]{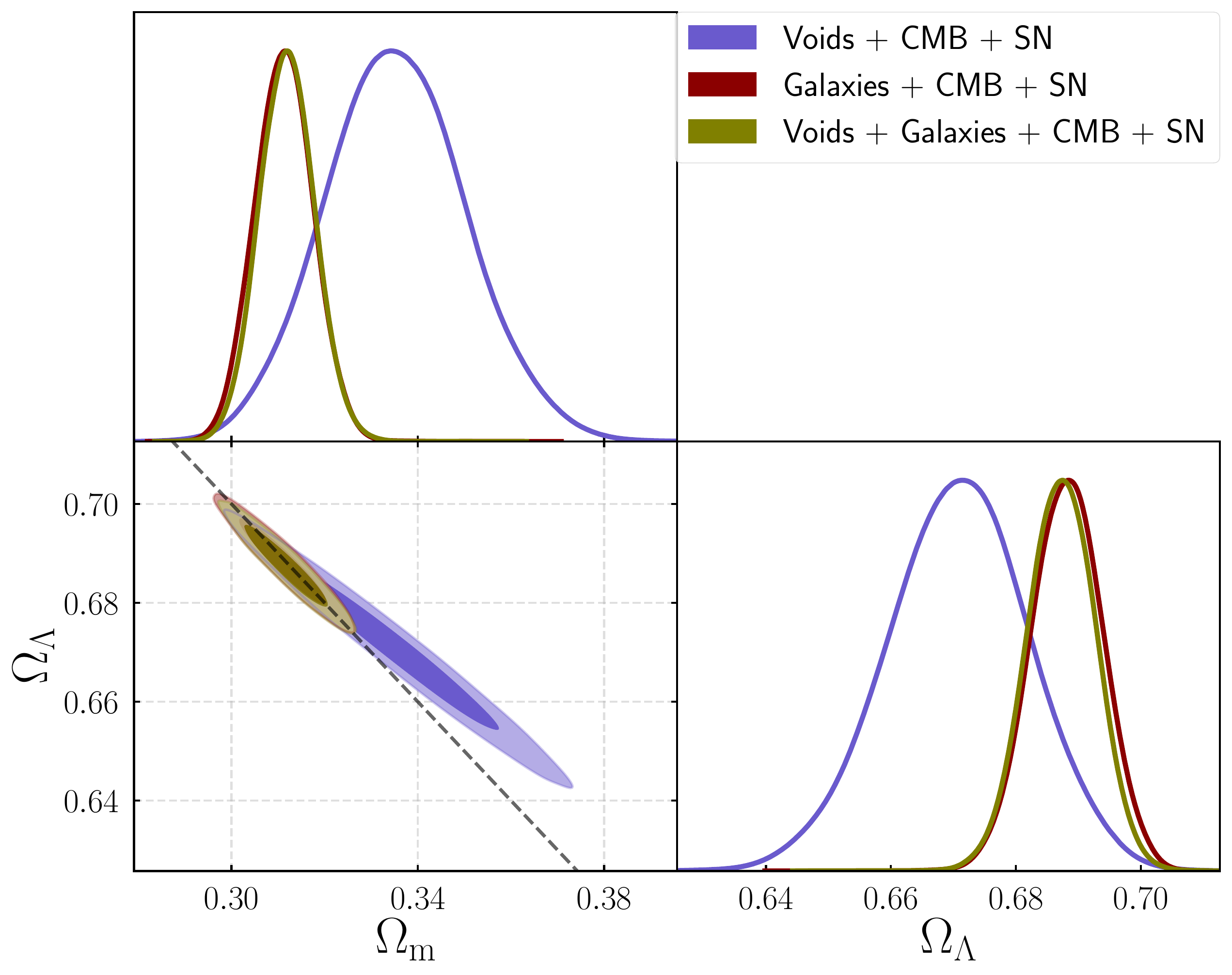}

\caption{Marginalized posterior constraints on cosmological parameters under the assumption of an $o$CDM cosmology. Shaded contours show the 68\% and 95\% confidence limit regions and corresponding mean values and marginalized 68\% credible intervals are shown in Table~\ref{table:resultantConstraints}. On the left, we show results from SDSS Voids and Galaxies alone and the combination of these two datasets.  CMB measurements from Planck TT,TE,EE+lowE+lensing and SN measurements from Pantheon SN Ia sample are combined with these three combinations to provide tighter constraints. The right plot shows only the combination of SDSS data with Planck. The addition of void information results in a significant gain of information both with and without the inclusion of CMB + SN data. The dashed line corresponds to the $\Lambda$CDM prediction of a spatially flat Universe with $\Omega_\mathrm{k}=0$.}
\label{fig:oCDM}
\end{figure*}

An uninformative, flat prior of $-1 < \Omega_\mathrm{k} < 1$ is imposed. Figure~\ref{fig:oCDM} shows constraints for this cosmology and Table~\ref{table:results} lists the mean values and marginalized 68\% credible intervals obtained for various combinations of measurement techniques. All models show a reasonable agreement with a $\Lambda$CDM prediction for a flat Universe.

Void constraints at a single redshift are unable to close contours in the $\Omega_\mathrm{m}-\Omega_\Lambda$ plane \citep{Woodfinden:2022,Nadathur:2020a}. This is because measurement of $D_\mathrm{M}/D_\mathrm{H}$ at a single redshift leads to a perfect degeneracy between these two parameters. Over a wide range in redshift, the variation in loci of models in the $\Omega_\mathrm{m}-\Omega_\Lambda$ plane that result in the same measurement of $D_\mathrm{M}/D_\mathrm{H}$ breaks this degeneracy (see Figure~\ref{fig:gradient}). In this work there is still a strong degeneracy between these parameters that is not broken enough to close the contours in the $\Omega_\mathrm{m}-\Omega_\Lambda$ plane; however, a narrowing can be observed. 

The combination of Planck and SN only measures $\Omega_m = 0.336\pm 0.022$, $\Omega_\mathrm{k} = -0.0062^{+0.0062}_{-0.0054}$ and $H_0 = 65.1\pm 2.2$. The additional information from SDSS (voids+galaxies) results in a $32\%$ reduction in the measured error on $\Omega_\mathrm{m}$, a $68\%$ reduction in the measured error on $\Omega_\mathrm{k}$ and a $71\%$ reduction in the measured error on $H_0$.

\subsection{owCDM}

We next consider an $ow$CDM model, where $\Omega_\mathrm{k}$ is allowed to vary so that we are not imposing a flat curvature and allow the constant equation of state parameter for dark energy $w_0$ to vary. In a $\Lambda$CDM model $\Omega_\mathrm{k} = 0$ and $w_0 = -1$.

\begin{figure*}
\centering
\includegraphics[width=.48\textwidth]{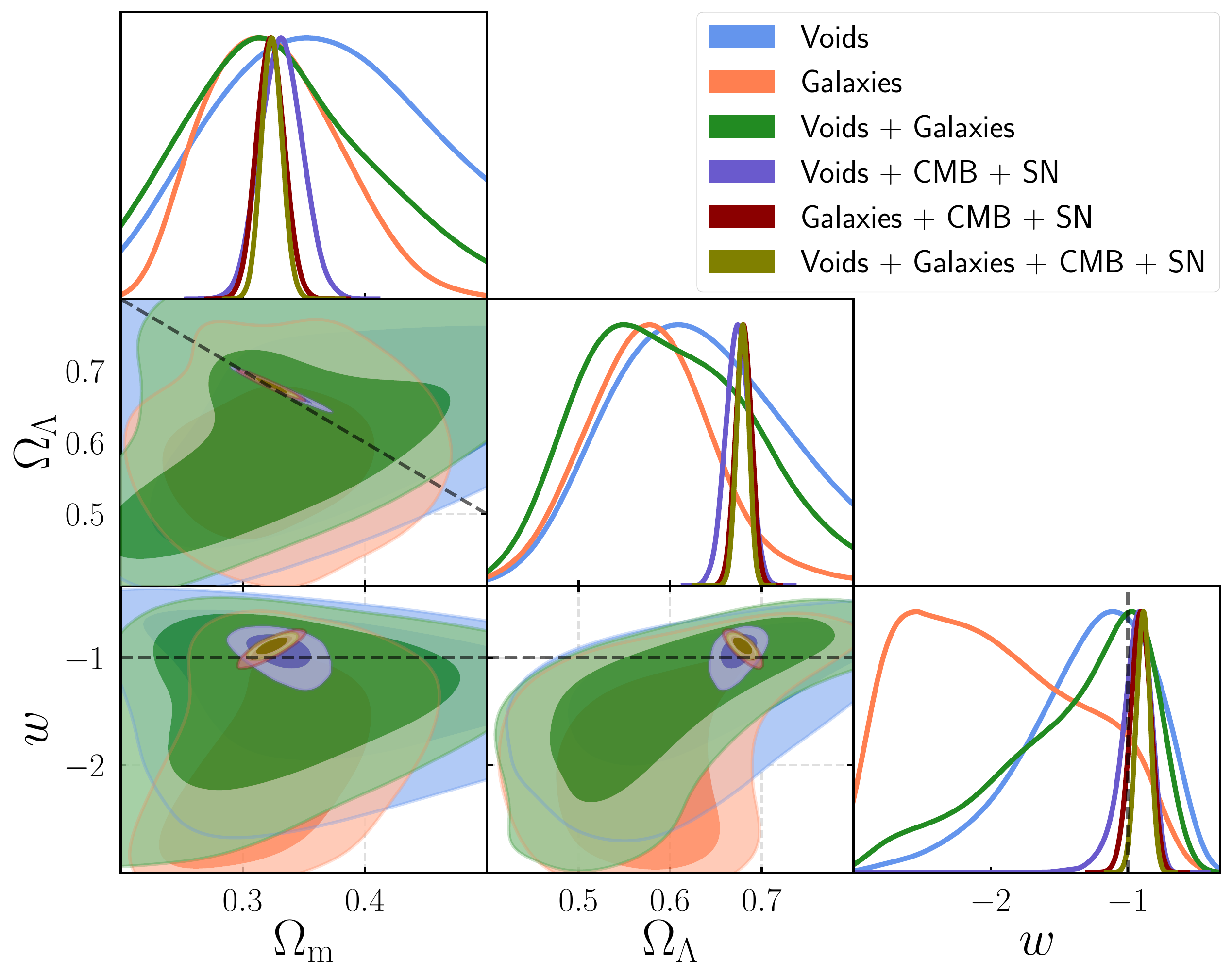}\quad
\includegraphics[width=.48\textwidth]{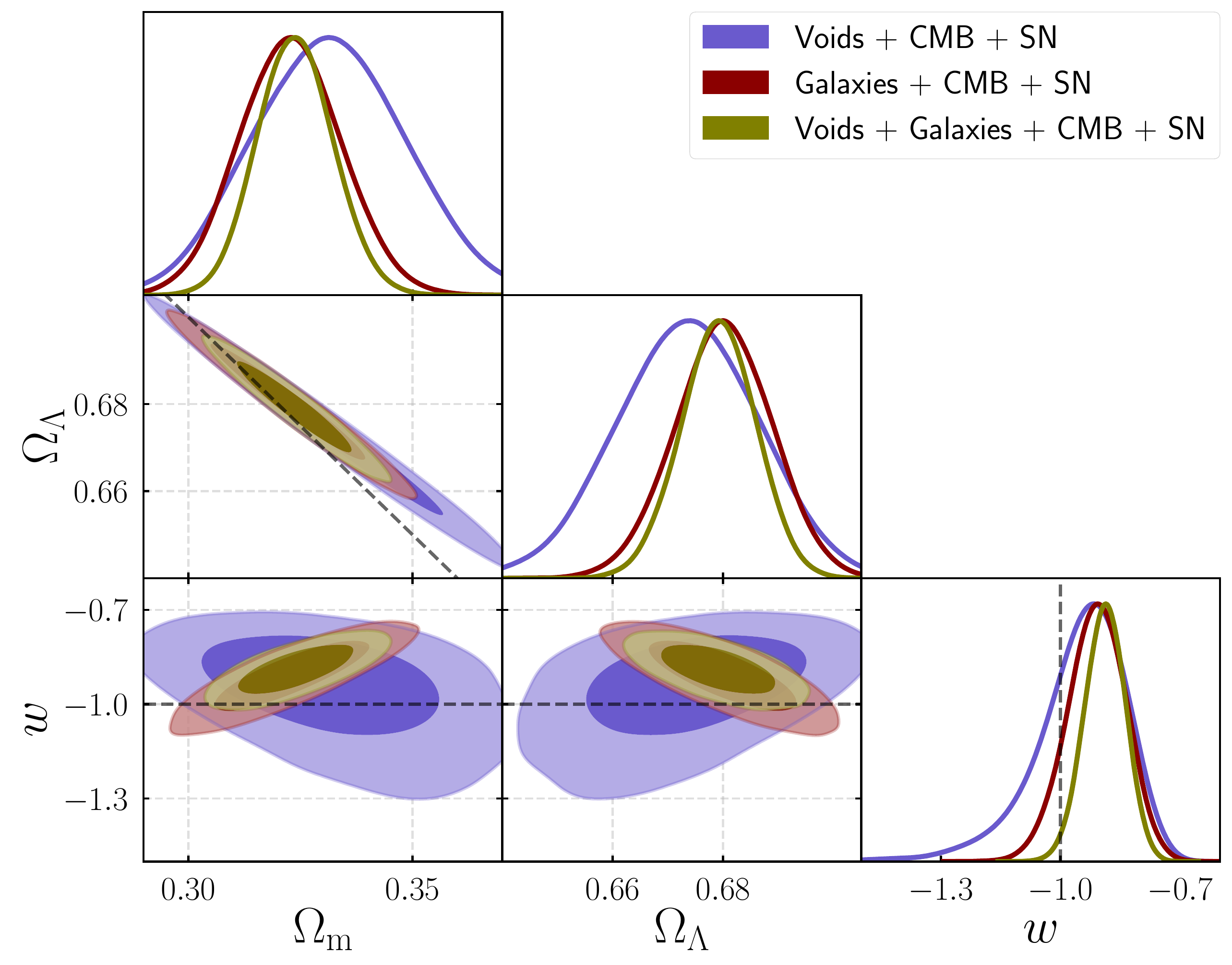}

\caption{Marginalized posterior constraints on cosmological parameters under the assumption of an $ow$CDM cosmology. Shaded contours show the 68\% and 95\% confidence limit regions and corresponding mean values and marginalized 68\% credible intervals are shown in Table~\ref{table:resultantConstraints}. On the left, we show results from SDSS Voids and Galaxies alone and their combination. Additionally, we show these three combinations with the addition of CMB measurements from Planck TT,TE,EE+lowE+lensing and SN measurements from the Pantheon SN Ia sample. The right plot shows only the combination of SDSS data with Planck. The addition of void information results in a significant gain of information both with and without the inclusion of CMB + SN data. The dashed line corresponds to the $\Lambda$CDM prediction of $w=-1$ and a spatially flat Universe with $\Omega_\mathrm{k}=0$.}
\label{fig:owCDM}
\end{figure*}

Uninformative, flat priors of $-3 < w < -1/3$ and $-1.0 < \Omega_\mathrm{k} < 1.0$ are imposed. Figure~\ref{fig:owCDM} shows results for this cosmology and Table~\ref{table:resultantConstraints} lists the mean values and marginalized 68\% credible intervals for a variety of combinations of different measurement techniques. We find reasonable agreement with flat spatial curvature with all values of $\Omega_\mathrm{k}$ consistent with $0$ and $w_0$ consistent with $-1$ within $2 \sigma$.

\subsection{ow(z)CDM}

We now consider to a $ow(z)$CDM model. We do not impose spatial flatness, allowing $\Omega_\mathrm{k}$ to be a free parameter. We also allow for a time-varying equation of state parameter $w$ as in Section~\ref{sec:wCDM}. In a $\Lambda$CDM model $\Omega_\mathrm{k}=0$, $w=-1$, and $w_\mathrm{a}=0$.

\begin{figure}
\centering
\includegraphics[width=.48\textwidth]{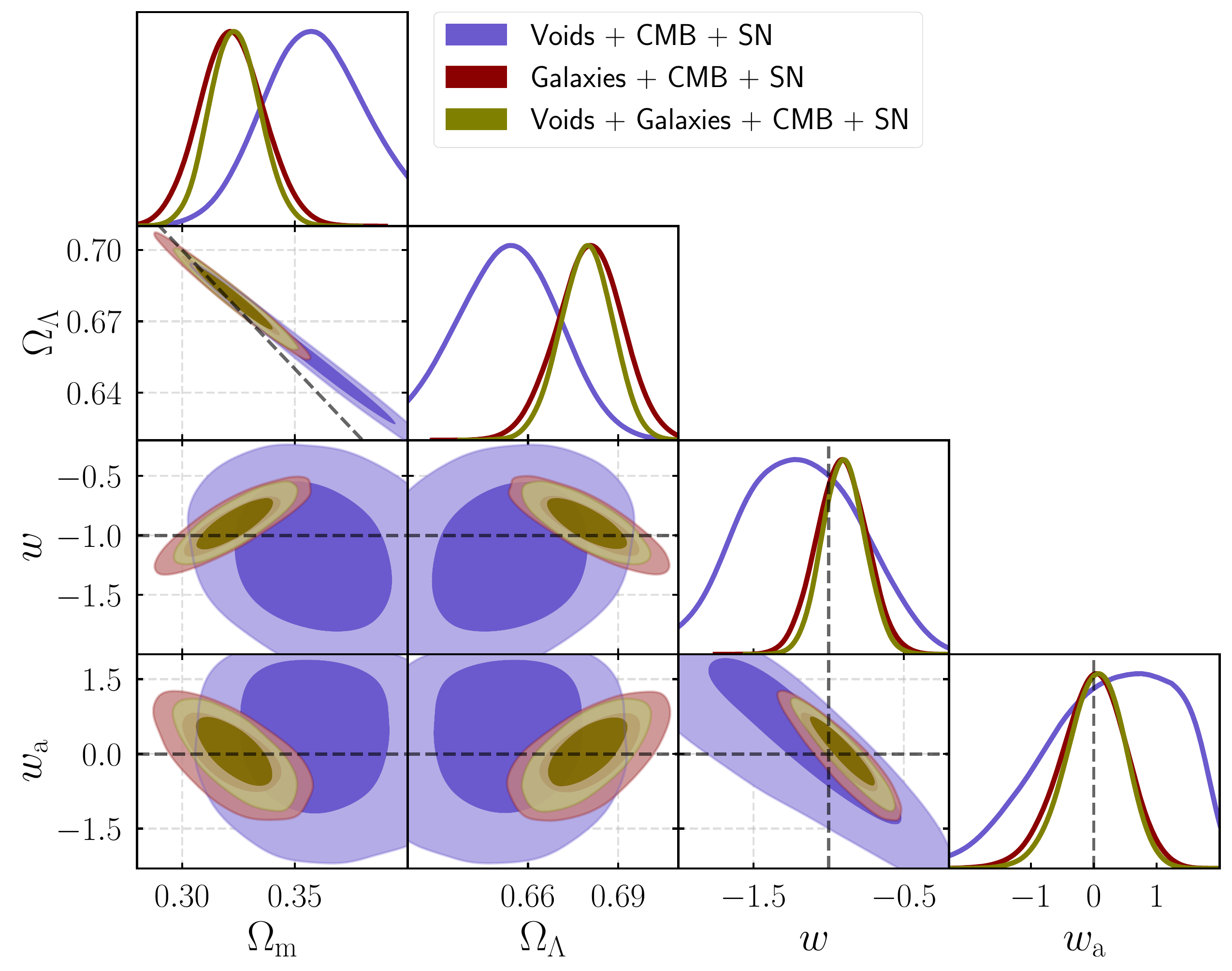}
\caption{Marginalized posterior constraints on cosmological parameters under the assumption of a $ow(z)$CDM cosmology. Shaded contours show the 68\% and 95\% confidence limit regions and corresponding mean values and marginalized 68\% credible intervals are shown in Table~\ref{table:resultantConstraints}. The $w_\mathrm{a}$ constraint for the Void + CMB + SN result can be seen to be hitting the upper prior boundary of $w_\mathrm{a} = 2$ and as such is not reported in our results. The dashed line corresponds to the $\Lambda$CDM prediction of $w=-1$, $w_\mathrm{a}=0$, and a spatially flat Universe with $\Omega_\mathrm{k}=0$.}
\label{fig:owwaCDM}
\end{figure}

Uninformative, flat priors of $-3 < w < -1/3$, $-3 < w_\mathrm{a} < 2$, and $-1 < \Omega_\mathrm{k} < 1$ are imposed. Figure~\ref{fig:owwaCDM} shows results for this cosmology and Table~\ref{table:resultantConstraints} lists the mean values and marginalized 68\% credible intervals.  Only the combination of SDSS data with CMB and SN results from \citet{Planck:2020} and \citet{Scolnic:2017} are shown, as SDSS alone does not have enough constraining power within a reasonable prior volume. Once again we see reasonable agreement with flat spatial curvature, $w_0 = -1$, and $w_\mathrm{a} = 0$ across all measurement techniques. For the combination of voids with CMB and SN data, the constraints on $w_\mathrm{a}$ can be seen to be hitting the upper bound of the prior; as such only lower limits are given in Table~\ref{table:results}.

\section{Conclusions}\label{sec:conclusions}

We have presented the cosmological implications of void-galaxy and galaxy-galaxy clustering over a wide redshift range in the SDSS DR7 (MGS), SDSS DR12 (BOSS), and SDSS DR16 (eBOSS) galaxy surveys. Void-galaxy and galaxy-galaxy clustering results are combined, taking into account cross-correlations between different measurement techniques and redshift bins. We see a significant gain of information from the combination of these measurement techniques, as well as from the inclusion of data from a wide range in redshift. The constraints on $D_\mathrm{M}/D_\mathrm{H}$ have a strong degeneracy following a locus of models that predict the same value (see Figure~\ref{fig:gradient}). The direction of this locus changes with redshift meaning that void measurements over a wide redshift range improve the cosmological constraints presented in this work beyond increased precision solely due to an increased volume measured. 

We perform a Bayesian analysis, comparing these measurements to various cosmological models including base $\Lambda$CDM, a constant dark energy equation of state allowed to vary from $w=-1$ ($w$CDM), a time-varying dark energy equation of state ($w(z)$CDM), allowing for spatial curvature ($oCDM$), allowing for spatial curvature with a constant dark energy equation of state allowed to vary from $w=-1$ ($ow$CDM), allowing for spatial curvature with a time-varying dark energy equation of state ($ow(z)$CDM). Final constraints on these cosmologies are shown in Table~\ref{table:results}. Results from SDSS have also been combined with CMB results from \citet{Planck:2020} and Pantheon SN Ia sample from \citet{Scolnic:2017} to provide extremely tight constraints on these cosmological models. With currently available data, parameter estimations and associated errors are heavily influenced by the already stringent CMB constraints. The inclusion of void information does improve the CMB constraints and is of increasing importance for models with more freedom in the low-redshift behaviour. This is true for many low-redshift cosmological constraints. However, a clear improvement can be seen in many parameter constraints from the inclusion of void data in addition to bringing in BAO \& RSD and SN measurements. In the near future, upcoming data from next-generation surveys will offer comparable uncertainty to Planck CMB results using void-galaxy and galaxy-galaxy clustering alone \citep[e.g.][]{DESI:2016,Radinovic:2023}.

The combination of void-galaxy and galaxy-galaxy clustering results in SDSS provides a large gain of information compared to that from galaxy-galaxy clustering alone. Results for $\Lambda$CDM provide a $43$\% reduction in the size of the error on $\Omega_\mathrm{m}$ compared to galaxy clustering alone. Results for $w$CDM provide a $44$\% and $45$\% reduction in the size of the error on $\Omega_\mathrm{m}$ and $w$. Results for $w(z)$CDM provide a $4$\% and $9$\%, and $16$\% reduction in the size of the error on $\Omega_\mathrm{m}$, $w$, and $w(z)$ when used in combination with CMB measurements from Planck and SN measurements from Pantheon. Results for $o$CDM provide a $35$\% and $11$\% reduction in the size of the error on $\Omega_\Lambda$ and $\Omega_\mathrm{k}$ compared to galaxy clustering alone. Results for $ow$CDM provide a $31$\%, and $38$\% reduction in the size of the error on $\Omega_\mathrm{m}$, and $w$ when used in combination with CMB measurements from Planck and SN from Pantheon. Finally, results for $ow(z)$CDM provide a $21$\%, $15$\%, $18$\%, and $13$\% reduction in the size of the error on $\Omega_\mathrm{m}$, $\Omega_\mathrm{k}$, $w_0$, and $w_\mathrm{a}$ when used in combination with CMB + SN measurements. 

Void-galaxy and galaxy-galaxy clustering results are also combined with CMB + SN results as shown in Table~\ref{table:results}. The addition of void-galaxy and galaxy-galaxy data can further increase the precision of cosmological parameters shown in various models. All results shown are in excellent agreement with a base $\Lambda$CDM cosmological model. 

Our work shows the importance of including voids as a cosmological probe and motivates the inclusion of voids as standard in the analysis of galaxy surveys due to the large information gain provided, especially when used in combination with measurements from other techniques. This is particularly important for low-redshift geometrical tests of cosmic expansion and discriminating between alternative cosmological models. Results expected shortly from DESI and Euclid will probe larger volumes over a wider redshift range. Void-galaxy and galaxy-galaxy clustering results in these surveys will provide powerful tests of the behaviour of cosmological models at low redshift.

\section*{Acknowledgments}

AW and WP acknowledge the support of the Natural Sciences and Engineering Research Council of Canada (NSERC) [funding reference number 547744 and RGPIN-2019-03908 respectively]. 
SN acknowledges support from an STFC Ernest Rutherford Fellowship, grant reference ST/T005009/2.

Research at Perimeter Institute is supported in part by the Government of Canada through the Department of Innovation, Science and Economic Development Canada and by the Province of Ontario through the Ministry of Colleges and Universities. This research was enabled in part by support provided by Compute Ontario (computeontario.ca) and Compute Canada (computecanada.ca).

Funding for the Sloan Digital Sky Survey IV has been provided by the Alfred P. Sloan Foundation, the U.S. Department of Energy Office of Science, and the Participating Institutions. 

SDSS-IV acknowledges support and resources from the Center for High Performance Computing at the University of Utah. The SDSS website is www.sdss.org.

SDSS-IV is managed by the Astrophysical Research Consortium for the Participating Institutions of the SDSS Collaboration including the Brazilian Participation Group, the Carnegie Institution for Science, 
Carnegie Mellon University, Center for Astrophysics | Harvard \& Smithsonian, the Chilean Participation 
Group, the French Participation Group, Instituto de Astrof\'isica de Canarias, The Johns Hopkins University, Kavli Institute for the Physics and Mathematics of the Universe (IPMU) / University of Tokyo, the Korean Participation Group, Lawrence Berkeley National Laboratory, Leibniz Institut f\"ur Astrophysik Potsdam (AIP),  Max-Planck-Institut f\"ur Astronomie (MPIA Heidelberg), Max-Planck-Institut f\"ur Astrophysik (MPA Garching), Max-Planck-Institut f\"ur Extraterrestrische Physik (MPE), National Astronomical Observatories of China, New Mexico State University, New York University, University of Notre Dame, Observat\'ario Nacional / MCTI, The Ohio State University, Pennsylvania State University, Shanghai Astronomical Observatory, United Kingdom Participation Group, Universidad Nacional Aut\'onoma 
de M\'exico, University of Arizona, University of Colorado Boulder, University of Oxford, University of Portsmouth, University of Utah, University of Virginia, University of Washington, University of Wisconsin, Vanderbilt University, and Yale University.

For the purpose of open access, the authors have applied a CC BY public copyright licence to any Author Accepted Manuscript version arising.

\section*{Data Availability}

Data supporting this research including the cross-covariance matrices and resulting likelihood for cosmological parameters will be made public after acceptance at \url{https://github.com/alexwoodfinden/SDSS-Void-Cosmology}. Data can also be obtained in advance of acceptance by reasonable
request to the corresponding author. 



\bibliographystyle{mnras}
\bibliography{paper}






\bsp	
\label{lastpage}
\end{document}